# Coalitional Game Theory for Communication Networks: A Tutorial

Walid Saad[1], Zhu Han[2], Mérouane Debbah[3], Are Hjørungnes[1] and Tamer Başar[4]

[1] UNIK - University Graduate Center, University of Oslo, Kjeller, Norway, email: {saad,arehj}@unik.no
[2] Electrical and Computer Engineering Department, University of Houston, Houston, USA, email: zhan2@mail.uh.edu
[3] Alcatel-Lucent Chair in Flexible Radio, SUPÉLEC, Gif-sur-Yvette, France, email: merouane.debbah@supelec.fr
[4] Coordinated Science Laboratory, University of Illinois at Urbana-Champaign, USA, email: basar1@illinois.edu

**Abstract**

Game theoretical techniques have recently become prevalent in many engineering applications, notably in communications. With the emergence of cooperation as a new communication paradigm, and the need for self-organizing, decentralized, and autonomic networks, it has become imperative to seek suitable game theoretical tools that allow to analyze and study the behavior and interactions of the nodes in future communication networks. In this context, this tutorial introduces the concepts of cooperative game theory, namely coalitional games, and their potential applications in communication and wireless networks. For this purpose, we classify coalitional games into three categories: Canonical coalitional games, coalition formation games, and coalitional graph games. This new classification represents an application-oriented approach for understanding and analyzing coalitional games. For each class of coalitional games, we present the fundamental components, introduce the key properties, mathematical techniques, and solution concepts, and describe the methodologies for applying these games in several applications drawn from the state-of-the-art research in communications. In a nutshell, this article constitutes a unified treatment of coalitional game theory tailored to the demands of communications and network engineers.

This work was done during the stay of Walid Saad at the Coordinated Science Laboratory, University of Illinois at Urbana-Champaign and was supported by the Research Council of Norway through projects 183311/S10, 176773/S10, and 18778/V11.

# I. INTRODUCTION AND MOTIVATION

Game theory provides a formal analytical framework with a set of mathematical tools to study the complex interactions among rational players. Throughout the past decades, game theory has made revolutionary impact on a large number of disciplines ranging from engineering, economics, political science, philosophy, or even psychology [1]. In recent years, there has been a significant growth in research activities that use game theory for analyzing communication networks. This is mainly due to: (i)- The need for developing autonomous, distributed, and flexible mobile networks where the network devices can make independent and rational strategic decisions; and (ii)- the need for low complexity distributed algorithms that can efficiently represent competitive or collaborative scenarios between network entities.

In general, game theory can be divided into two branches: non-cooperative [2] and cooperative game theory [1], [3]. *Non-cooperative game theory* studies the strategic choices resulting from the interactions among *competing* players, where each player chooses its strategy independently for improving its own performance (utility) or reducing its losses (costs). For solving non-cooperative games, several concepts exist such as the celebrated Nash equilibrium [2]. The mainstream of existing research in communication networks focused on using non-cooperative games in various applications such as distributed resource allocation [4], congestion control [5], power control [6], and spectrum sharing in cognitive radio, among others. This need for non-cooperative games led to numerous tutorials and books outlining its concepts and usage in communication, e.g., [7], [8].

While non-cooperative game theory studies competitive scenarios, *cooperative game theory* provides analytical tools to study the behavior of rational players *when they cooperate*. The main branch of cooperative games describes the formation of cooperating groups of players, referred to as coalitions [1], that can strengthen the players' positions in a game. In this tutorial, we restrict our attention to coalitional game theory albeit some other references can include other types of games, such as bargaining, under the umbrella of cooperative games. Coalitional games have also been widely explored in different disciplines such as economics or political science. Recently, cooperation has emerged as a new networking paradigm that has a dramatic effect of improving the performance from the physical layer [9], [10] up to the networking layers [4]. However, implementing cooperation in large scale communication networks faces several challenges such as adequate modeling, efficiency, complexity, and fairness, among others. Coalitional games prove to be a very powerful tool for designing fair, robust, practical, and efficient cooperation strategies in communication networks. Most of the current research in the field is restricted to applying standard coalitional game models and techniques to study very limited aspects of cooperation in networks. This is mainly due to the sparsity of the literature that tackles coalitional games. In fact, most pioneering game theoretical references, such as [1–3], focus on non-cooperative games; touching slightly on coalitional games within a few chapters.

In this article, we aim to provide a unified treatment of coalitional game theory oriented towards engineering applications. Thus, the goal is to gather the state-of-the-art research contributions, from game theory and communications, that address the major opportunities and challenges in applying coalitional games to the understanding and designing of modern communication systems, with emphasis on both new analytical techniques and novel application scenarios. With the incessant growth in research revolving around cooperation, self-organization and fairness in communication networks, this tutorial constitutes a comprehensive guide that enables to fully exploit the potential of coalitional game theory. The tutorial starts by laying out the main components of coalitional games in Section II while in the following sections it zooms in on an in-depth study of these games and their applications. Since the literature on coalitional games and their communication applications is sparse, we introduce a novel classification of coalitional games which allows grouping of various types of games under one class based on several game properties. Hence, we group coalitional games into three distinct classes:

1) **Class I:** Canonical (coalitional) games[1]
2) **Class II:** Coalition formation games
3) **Class III:** Coalitional graph games

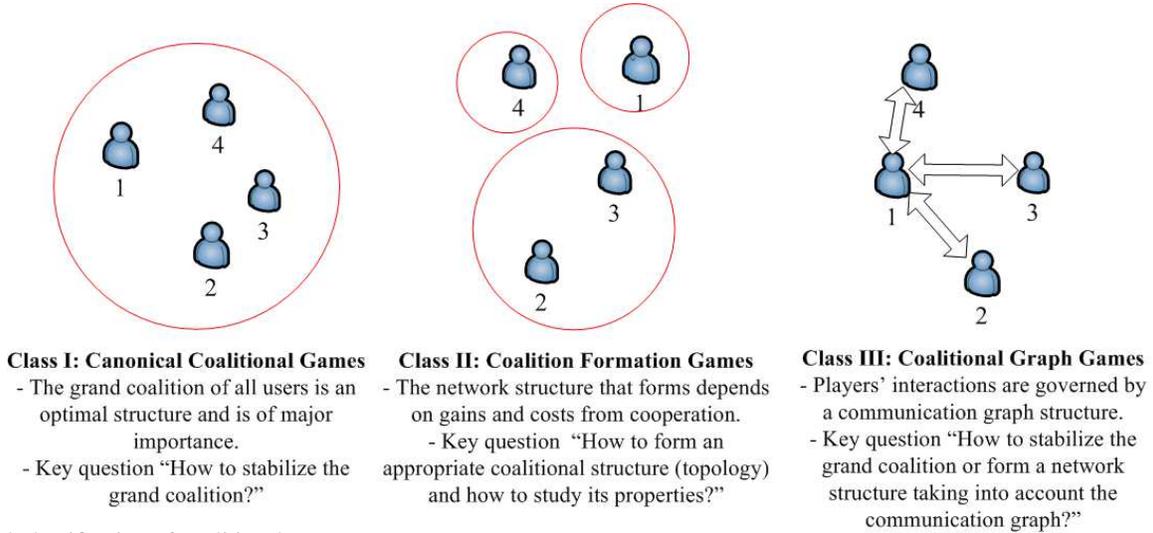

Fig. 1. A novel classification of coalitional games.

This novel classification is intended to provide an application-oriented approach to coalitional games. The key features of these classes are summarized in Fig. 1 and an in-depth study of each class is provided in Sections III, IV, and V.

## II. COALITIONAL GAME THEORY: PRELIMINARIES

In essence, coalitional games involve a set of players, denoted by $\mathcal{N} = \{1, \ldots, N\}$ who seek to form cooperative groups, i.e., coalitions, in order to strengthen their positions in the game. Any coalition $S \subseteq \mathcal{N}$ represents an agreement between the players in $S$ to act as a single entity. The formation of coalitions or alliances is ubiquitous in many applications. For example, in political games, parties, or individuals can form coalitions for improving their voting power. In addition to the player set $\mathcal{N}$, the second fundamental concept of a coalitional game is the coalition *value*. Mainly, the coalition value, denoted by $v$, quantifies the worth of a coalition in a game. The definition of the coalition value determines the *form* and *type* of the game. Nonetheless, independent of the definition of the value, a coalitional game is uniquely defined by the pair $(\mathcal{N}, v)$. It must be noted that the value $v$ is, in many instances, referred to as *the game*, since for every $v$ a different game may be defined.

The most common form of a coalitional game is the *characteristic form*, whereby the value of a coalition $S$ depends *solely* on the members of that coalition, with no dependence on how the players in $\mathcal{N} \setminus S$ are structured. The characteristic form was introduced, along with a category of coalitional games known as games with *transferable utility* (TU), by Von Neuman and Morgenstern [11]. The value of a game in characteristic form with TU is a function over the real line defined as $v : 2^N \to \mathbb{R}$ (characteristic function). This characteristic function associates with every coalition $S \subseteq \mathcal{N}$ a real number quantifying the gains of $S$. The TU property implies that the total utility represented by this real number can be divided in any manner between the coalition members. The values in TU games are thought of as monetary values that the members in a coalition can distribute among themselves using an appropriate *fairness* rule (one such rule being an equal distribution of the utility). The amount of utility that a player $i \in S$ receives from the division of $v(S)$ constitutes the player's *payoff* and is denoted by $x_i$ hereafter. The vector $\boldsymbol{x} \in \mathbb{R}^{|S|}$ ($|\cdot|$ represents the cardinality of a set) with each element $x_i$ being the payoff of player $i \in S$ constitutes a *payoff allocation*. Although the TU characteristic function can model a broad range of games, many scenarios exist where the coalition value cannot be assigned a single real number, or rigid restrictions exist on the distribution of the utility. These games are known as *coalitional games with non-transferable utility (NTU)* and were first introduced by Aumann and Peleg using non-cooperative strategic games as a basis [1], [12]. In an NTU game, the payoff that each player in a coalition $S$ receives is dependent on the joint actions that the players of coalition $S$ select[2]. The value of a coalition $S$ in an NTU game, $v(S)$, is no longer a function over the real line, but a set of payoff vectors, $v(S) \subseteq \mathbb{R}^{|S|}$, where each element $x_i$ of a vector $\boldsymbol{x} \in v(S)$ represents a payoff that player $i \in S$ can obtain within coalition $S$ given a certain strategy selected by $i$ while being a member of $S$. Given this definition, a TU game can be seen as a particular case of the NTU framework [1]. Coalitional games in characteristic form with TU or NTU constitute one of the most important types of games, and their solutions are explored in detail in the following sections.

---
[1] We will use the terminologies "canonical coalitional games" and "canonical games" interchangeably throughout this tutorial.

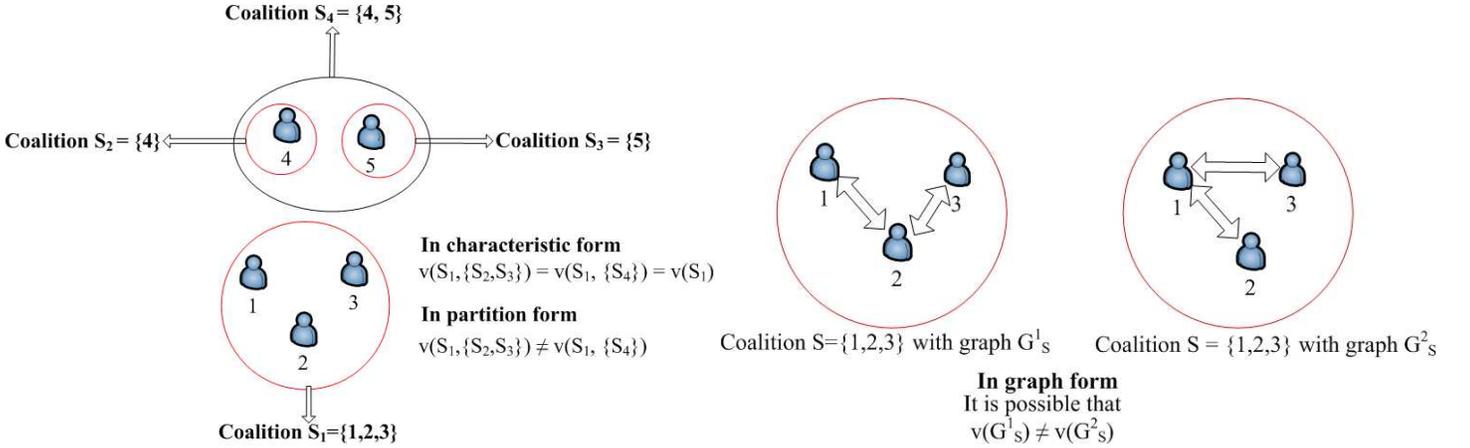

Fig. 2. (a) Coalitional games in characteristic form vs. partition form.  (b) Example of a coalitional game in graph form.

Recently, there has been an increasing interest in coalitional games where the value of a coalition depends on the partition of $\mathcal{N}$ that is in place at any time during the game. In such games, unlike the characteristic form, the value of a coalition $S$ will have a strong dependence on how the players in $\mathcal{N} \setminus S$ are structured. For this purpose, Thrall and Lucas [13] introduced the concept of games in *partition form*. In these games, given a *coalitional structure* $\mathcal{B}$, defined as a *partition* of $\mathcal{N}$, i.e., a collection of coalitions $\mathcal{B} = \{B_1, \ldots, B_l\}$, such that $\forall\, i \neq j,\ B_i \cap B_j = \emptyset$, and $\cup_{i=1}^{l} B_i = \mathcal{N}$, the value of a coalition $S \in \mathcal{B}$ is defined as $v(S, \mathcal{B})$. This definition imposes a dependence on the coalitional structure when evaluating the value of $S$. Coalitional games in partition form are inherently complex to solve; however, the potential of these games is interesting and, thus, we will provide insights on these games in the following sections.

As an example on the difference between characteristic and partition forms, consider a 5-players game with $\mathcal{N} = \{1,2,3,4,5\}$ and let $S_1 = \{1,2,3\}$, $S_2 = \{4\}$, $S_3 = \{5\}$, and $S_4 = \{4,5\}$. Given two partitions $\mathcal{B}_1 = \{S_1, S_2, S_3\}$ and $\mathcal{B}_2 = \{S_1, S_4\}$ of $\mathcal{N}$, evaluating the value of coalition $S_1$ depends on the form of the game. If the game is in *characteristic form*, then $v(S_1, \mathcal{B}_1) = v(S_1, \mathcal{B}_2) = v(S_1)$ while in *partition form* $v(S_1, \mathcal{B}_1) \neq v(S_1, \mathcal{B}_2)$ (the value here can be either TU or NTU). The basic difference is that, unlike the characteristic form, the value of $S_1$ in partition form depends on whether players $4$ and $5$ cooperate or not. This is illustrated in Fig. 2 (a).

---

[2]The action space depends on the underlying non-cooperative game (see [12] for examples).

In many coalitional games, the players are interconnected and communicate through pairwise links in a graph. In such scenarios, both the characteristic form and the partition form may be unsuitable since, in both forms, the value of a coalition $S$ is independent of how the members of $S$ are connected. For modeling the interconnection graphs, coalitional games in *graph form* were introduced by Myerson in [14] where *connected* graphs were mapped into coalitions. This work was generalized in [15] by making the value of each coalition $S \subseteq \mathcal{N}$ a function of the graph structure connecting the members of $S$. Hence, given a coalitional game $(\mathcal{N}, v)$ and a graph $G_S$ (directed or undirected) with vertices the members of a coalition $S \subseteq \mathcal{N}$, the value of $S$ in *graph form* is given by $v(G_S)$. For games in graph form, the value can also depend on the graph $G_{\mathcal{N} \setminus S}$ interconnecting the players in $\mathcal{N} \setminus S$. An example of a coalitional game in graph form is given in Fig. 2 (b). In this figure, given two graphs $G_S^1 = \{(1,2), (2,3)\}$ and $G_S^2 = \{(1,2), (1,3)\}$ (a pair $(i,j)$ is a link between two players $i$ and $j$) defined over coalition $S = \{1, 2, 3\}$, a coalitional game in graph form could assign a different value for coalition $S$ depending on the graph[3]. Hence, in graph form, it is possible that $v(G_S^1) \neq v(G_S^2)$, while in characteristic or partition form, the presence of the graph does not affect the value. Having introduced the fundamental concepts for coalitional games, the rest of this tutorial provides an in-depth analysis of each class of games.

## III. CLASS I: CANONICAL COALITIONAL GAMES

### A. Main Properties of Canonical Coalitional Games

Under the class of canonical coalitional games, we group the most popular category of games in coalitional game theory. Hence, this class pertains to the coalitional games tools that have been widely understood, thoroughly formalized, and have clear solution concepts. For classifying a game as canonical, the main requirements are as follows:

1) The coalitional game is in *characteristic* form (TU or NTU).
2) Cooperation, i.e., the formation of large coalitions, is *never* detrimental to any of the involved players. Hence, in canonical games no group of players can do worse by cooperating, i.e., by joining a coalition, than by acting non-cooperatively. This pertains to the mathematical property of *superadditivity*.
3) The main objectives of a canonical game are: (i)- To study the properties and stability of the *grand coalition*, i.e., the coalition of all the players in the game, and (ii)- to study the gains resulting from cooperation with negligible or no cost, as well as the distribution of these gains in a *fair* manner to the players.

The first two conditions for classifying a game as canonical pertain to the mathematical properties of the game. First, any canonical game must be in characteristic form. Second, the canonical game must be superadditive, which is defined as

$$v(S_1 \cup S_2) \supset \{\boldsymbol{x} \in \mathbb{R}^{|S_1 \cup S_2|} | (x_i)_{i \in S_1} \in v(S_1), (x_j)_{j \in S_2} \in v(S_2)\} \ \forall S_1 \subset \mathcal{N}, S_2 \subset \mathcal{N}, \ \text{s.t.} \ S_1 \cap S_2 = \emptyset, \quad (1)$$

where $\boldsymbol{x}$ is a payoff allocation for coalition $S_1 \cup S_2$. Superadditivity implies that, given any two disjoint coalitions $S_1$ and $S_2$, if coalition $S_1 \cup S_2$ forms, then it can give its members any allocations they can achieve when acting in $S_1$ and $S_2$ separately. The definition in (1) is used in an NTU case. For a TU game, superadditivity reduces to [1]

$$v(S_1 \cup S_2) \geq v(S_1) + v(S_2) \ \forall S_1 \subset \mathcal{N}, S_2 \subset \mathcal{N}, \ \text{s.t.} \ S_1 \cap S_2 = \emptyset. \quad (2)$$

From (2), the concept of a superadditive game is better grasped. Simply, a game is superadditive if cooperation, i.e., the formation of a large coalition out of disjoint coalitions, guarantees at least the value that is obtained by the disjoint coalitions separately. The rationale behind superadditivity is that, within a coalition, the players can always revert back to their non-cooperative behavior to obtain their non-cooperative payoffs. Thus, in a superadditive game, cooperation is always beneficial. Due to superadditvity in canonical games, it is to the joint benefit of the players to always form the *grand coalition* $\mathcal{N}$, i.e, the coalition of *all* the players, since the payoff received from $v(\mathcal{N})$ is at least as large as the amount received by the players in any disjoint set of coalitions they could form. The formation of the grand coalition in canonical games implies that the main emphasis is on studying the properties of this grand coalition. Two key aspects are of importance in canonical games: (i)- Finding a payoff allocation which guarantees that no group of players have an incentive to leave the grand coalition (having a *stable* grand coalition), and (ii)- assessing the gains that the grand coalition can achieve as well as the fairness criteria that must be used for distributing these gains (having a *fair* grand coalition). For solving canonical coalitional games, the literature presents a number of concepts [1], [3] that we will explore in detail in the following sections.

---

[3] In this example we considered an undirected graph and a single link between every pair of nodes. However, multiple links between pairs of nodes as well as directed graphs can also be considered within the graph form of coalitional games.

*B. The Core as a Solution for Canonical Coalitional Games*

*1) Definition:* The most renowned solution concept for coalitional games, and for games classified as canonical in particular, is *the core* [1], [3]. The core of a canonical game is directly related to the grand coalition's stability. In a canonical coalitional game $(\mathcal{N}, v)$, due to superadditvity, the players have an incentive to form the grand coalition $\mathcal{N}$. Thus, the core of a canonical game is the set of payoff allocations which guarantees that no group of players has an incentive to leave $\mathcal{N}$ in order to form another coalition $S \subset \mathcal{N}$. For a TU game, given the grand coalition $\mathcal{N}$, a payoff vector $\boldsymbol{x} \in \mathbb{R}^N$ ($N = |\mathcal{N}|$) for dividing $v(\mathcal{N})$ is *group rational* if $\sum_{i \in \mathcal{N}} x_i = v(\mathcal{N})$. A payoff vector $\boldsymbol{x}$ is *individually rational* if every player can obtain a benefit no less than acting alone, i.e. $x_i \geq v(\{i\}), \forall\ i$. An *imputation* is a payoff vector satisfying the above two conditions. Having defined an imputation, *the core* is defined as

$$\mathcal{C}_{\text{TU}} = \left\{ \boldsymbol{x} : \sum_{i \in \mathcal{N}} x_i = v(\mathcal{N}) \text{ and } \sum_{i \in S} x_i \geq v(S)\ \forall\ S \subseteq \mathcal{N} \right\}. \tag{3}$$

In other words, the core is the set of imputations where no coalition $S \subset \mathcal{N}$ has an incentive to reject the proposed payoff allocation, deviate from the grand coalition and form coalition $S$ instead. The core guarantees that these deviations do not occur through the fact that any payoff allocation $\boldsymbol{x}$ that is in the core guarantees at least an amount of utility equal to $v(S)$ for every $S \subset \mathcal{N}$. Clearly, whenever one is able to find a payoff allocation that lies in the core, then the grand coalition is a stable and optimal solution for the coalitional game. For solving NTU games using the core, the value $v$ of the NTU game is often assumed to satisfy the following, for any coalition $S$, [1]: (1)- The value $v(S)$ of any coalition $S$ must be a closed and convex subset of $\mathbb{R}^{|S|}$, (2)- the value $v(S)$ must be *comprehensive*, i.e., if $\boldsymbol{x} \in v(S)$ and $\boldsymbol{y} \in \mathbb{R}^{|S|}$ are such that $\boldsymbol{y} \leq \boldsymbol{x}$, then $\boldsymbol{y} \in v(S)$, and (3)- the set $\{\boldsymbol{x} | \boldsymbol{x} \in v(S) \text{ and } x_i \geq z_i,\ \forall i \in S\}$ with $z_i = \max\{y_i | \boldsymbol{y} \in v(\{i\})\} < \infty\ \forall i \in \mathcal{N}$ must be a bounded subset of $\mathbb{R}^{|S|}$. The comprehensive property implies that if a certain payoff allocation $\boldsymbol{x}$ is achievable by the members of a coalition $S$, then, by changing their strategies, the members of $S$ can achieve any allocation $\boldsymbol{y}$ where $\boldsymbol{y} \leq \boldsymbol{x}$. The last property implies that, for a coalition $S$, the set of vectors in $v(S)$ in which each player in $S$ receives no less than the maximum that it can obtain non-cooperatively, i.e., $z_i$, is a bounded set. For a canonical NTU game $(\mathcal{N}, v)$ with $v$ satisfying the above properties, the core is defined as

$$\mathcal{C}_{\text{NTU}} = \{\boldsymbol{x} \in v(\mathcal{N}) | \forall S, \nexists \boldsymbol{y} \in v(S),\ \text{s.t.}\ y_i > x_i,\ \forall i \in S\}. \tag{4}$$

This definition for NTU also guarantees a stable grand coalition. The basic idea is that any payoff allocation in the core of an NTU game guarantees that no coalition $S$ can leave the grand coalition and provide a better allocation *for all* of its members. The difference from the TU case is that, in the NTU core, the grand coalition's stability is acquired over the elements of the payoff vectors while in the TU game, it is acquired by the sum of the payoff vectors' elements.

TABLE I
APPROACHES FOR FINDING THE CORE OF A CANONICAL COALITIONAL GAME

| |
|---|
| **Game theoretical and mathematical approaches** |
| (T1) - A *graphical* approach can be used for finding the core of TU games with up to 3 players. |
| (T2) - Using duality theory, a necessary and sufficient condition for the non-emptiness of the core exists through the *Bondareva-Shapley theorem* (Theorem 1) for TU and NTU [1], [3] . |
| (T3) - A class of canonical games, known as *convex coalitional games* always has a non-empty core. |
| (T4) - A necessary and sufficient condition for a non-empty core exists for a class of canonical games known as *simple games*, i.e., games where $v(S) \in \{0,1\}$, $\forall S \subseteq \mathcal{N}$ and $v(\mathcal{N}) = 1$. |
| **Application-specific approaches** |
| (T5) - In several applications, it suffices to find whether payoff distributions that are of interest in a given game, e.g., fair distributions, lie in the core. |
| (T6)- In many games, exploiting game-specific features such as the value's mathematical definition or the underlying nature and properties of the game model, helps finding the imputations that lie in the core. |

*2) Properties and Existence:* The cores of TU or NTU canonical games are not always guaranteed to exist. In fact, in many games, the core is empty and hence, the grand coalition cannot be stabilized. In these cases, alternative solution concepts may be used, as we will see in the following sections. However, coalitional game theory provides several categories of games which fit under our canonical game class, where the core is guaranteed to be non-empty. Before surveying the existence results for the core, we provide a simple example of the core in a TU canonical game:

*Example 1:* Consider a majority voting TU game $(\mathcal{N}, v)$ where $\mathcal{N} = \{1,2,3\}$. *The players, on their own, have no voting power, hence* $v(\{1\}) = v(\{2\}) = v(\{3\}) = 0$. *Any 2-players coalition wins two thirds of the voting power, and hence,* $v(\{1,2\}) = v(\{1,3\}) = v(\{2,3\}) = \frac{2}{3}$. *The grand coalition wins the whole voting power, and thus* $v(\{1,2,3\}) = 1$. *Clearly, this game is superadditive and is in characteristic form and thus is classified as canonical. By (3), solving the following inequalities yields the core and shows what allocations stabilize the grand coalition.*

$$x_1 + x_2 + x_3 = v(\{1,2,3\}) = 1, \ x_1 \geq 0, \ x_2 \geq 0, \ x_3 \geq 0,$$
$$x_1 + x_2 \geq v(\{1,2\}) = \frac{2}{3}, \ x_1 + x_3 \geq v(\{1,3\}) = \frac{2}{3}, \ x_2 + x_3 \geq v(\{2,3\}) = \frac{2}{3}.$$

*By manipulating these inequalities, the core of this game is found to be the* unique *vector* $\boldsymbol{x} = [\frac{1}{3} \ \frac{1}{3} \ \frac{1}{3}]$ *which corresponds of an equal division of the total utility of the grand coalition among all three players.*

In general, given a TU coalitional game $(\mathcal{N}, v)$ and an imputation $\boldsymbol{x} \in \mathbb{R}^N$, the core is found by a linear program (LP)

$$\min_{\boldsymbol{x}} \sum_{i \in \mathcal{N}} x_i, \ \text{s.t.} \ \sum_{i \in S} x_i \geq v(S), \ \forall S \subseteq \mathcal{N}. \tag{5}$$

The existence of the TU core is related to the feasibility of the LP in (5). In general, finding whether the core is non-empty through this LP, is NP-complete [16] due to the number of constraints growing exponentially with the number of players $N$ (this is also true for NTU games, see [1, Ch. 9.7]). However, for determining the non-emptiness of the core as well as finding the allocations that lie in the core several techniques exist and are summarized in Table I.

The *first* technique in Table I deals with TU games with up to 3 players. In such games, the core can be found using an easy graphical approach. The main idea is to plot the constraints of (5) in the plane $\sum_{i=1}^{3} x_i = v(\{1,2,3\})$. By doing so, the region containing the core allocation can be easily identified. Several examples on the graphical techniques are found in [3] and the technique for solving them is straightforward. Although the graphical method can provide a lot of intuition into the core of a canonical game, its use is limited to TU games with up to 3 players.

The *second* technique in Table I utilizes the dual of the LP in (5) to show that the core is non-empty. The main result is given through the Bondareva-Shapley theorem [1], [3] which relies on the *balanced* property. A TU game is *balanced* if and only if the inequality [1]

$$\sum_{S \subseteq \mathcal{N}} \mu(S) v(S) \leq v(\mathcal{N}), \tag{6}$$

is satisfied for all non-negative weight collections $\mu = (\mu(S))_{S \subseteq \mathcal{N}}$ ($\mu$ is a collection of weights, i.e., numbers in $[0,1]$, associated with each coalition $S \subseteq \mathcal{N}$) which satisfy $\sum_{S \supseteq i} \mu(S) = 1, \ \forall i \in \mathcal{N}$; this set of non-negative weights is known as a balanced set. This notion of a balanced game is interpreted as follows. Each player $i \in \mathcal{N}$ possesses a single unit of time, which can be distributed between all the coalitions that $i$ can be a member of. Every coalition $S \subseteq \mathcal{N}$ is active during a fraction of time $\mu(S)$ if all of its members are active during that time, and this coalition achieves a payoff of $\mu(S)v(S)$. In this context, the condition $\sum_{S \supseteq i} \mu(S) = 1$, $\forall i \in \mathcal{N}$ is simply a feasibility constraint on the players' time allocation, and the game is balanced if there is no feasible allocation of time which can yield a total payoff for the players that exceeds the value of the grand coalition $v(\mathcal{N})$. For NTU canonical games, an analogous definition for balancedness is found in [1], [3]. The definition for NTU is modified to accommodate the fact that the value $v$ in an NTU game is a set and not a function. Subsequently, given a TU or NTU balanced canonical game, the following result holds [1], [3].

**Theorem 1:** (Bondareva-Shapley) The core of a game is non-empty if and only if the game is balanced. ⋄

Therefore, in a given canonical game, one can always show that the core is non-empty by proving that the game is balanced through (6) for TU games or its counterpart for NTU [1, Ch. 9.7]. Proving the non-emptiness of the core through the balanced property is a popular approach and several examples on balanced games exist in the game theory literature [1], [3] as well as in the literature on communication networks [17], [18].

The *third* technique in Table I pertains to *convex* games. A TU canonical game is convex if
$$v(S_1) + v(S_2) \leq v(S_1 \cup S_2) + v(S_1 \cap S_2) \ \forall \ S_1, S_2 \subseteq \mathcal{N} \tag{7}$$

This convexity property implies that the value function, i.e., the game, is supermodular. Alternatively, a convex coalitional game is defined as any coalitional game that satisfies $v(S_1 \cup \{i\}) - v(S_1) \leq v(S_2 \cup \{i\}) - v(S_2)$, whenever $S_1 \subseteq S_2 \subseteq \mathcal{N} \setminus \{i\}$. This alternative definition implies that a game is convex if and only if for each player $i \in \mathcal{N}$ the marginal contribution of this player, i.e. the difference between the value of a coalition with and without this player, is nondecreasing with respect to set inclusion. The convexity property can also be extended to NTU in several ways, and the reader is referred to [3, Ch. 9.9] for more details. For both TU and NTU canonical games, a convex game is balanced and *has a non-empty* core, but the converse is not always true [3]. Thus, convex games constitute an important class of games where the core is non-empty. Examples of such games are ubiquitous in both game theory [1], [3] and communications [17].

The *fourth* technique pertains to *simple games* which are an interesting class of canonical games where the core can be shown to be non-empty. A simple game is a coalitional game where the value are either 0 or 1, i.e., $v(S) \in \{0, 1\}$, $\forall S \subseteq \mathcal{N}$ and the grand coalition has $v(\mathcal{N}) = 1$. These games model numerous scenarios, notably voting games. It is known that a simple game which contains at least one *veto* player $i \in \mathcal{N}$, i.e. a player $i$ such that $v(\mathcal{N} \setminus i) = 0$ has a *non-empty core* [3]. Moreover, in such simple games, the core is fully characterized, and it consists of all non-negative payoff profiles $\boldsymbol{x} \in \mathbb{R}^N$ such that $x_i = 0$ for each player $i$ that is a *non-veto* player, and $\sum_{i \in \mathcal{N}} x_i = v(\mathcal{N}) = 1$

The first four techniques in Table I rely mainly on well-known game theoretical properties. In many practical scenarios, notably in wireless and communication networking applications, alternative techniques may be needed to find the allocations in the core. These alternatives are inherently application-specific, and depend on the nature of the defined game and the properties of the defined value function. One of these alternatives, the *fifth* technique in Table I, is to investigate whether well-known allocation rules yield vectors that lie in the core. In many communication applications (and even game theoretical settings), the objective is to assess whether certain well-defined types of fair allocations such as equal fairness or proportional fairness among others are in the core or not, without finding all the allocations that are in the core. In such games, showing the non-emptiness of the core is done by testing whether such well-known allocations lie in the core or not, using the intrinsic properties of the considered game and using (3) for TU games or (4) for NTU games. A simple example of such a technique is Example 1, where one can check the non-emptiness of the core by easily showing that the equal allocation lies in the core. In many canonical games, the nature of the defined value for the game can be explored for showing the non-emptiness of the core; this is done in many applications such as in [10] where information theoretical properties are used, in [19] where network properties are used, as well as in [18], [20] where the value is given as a convex optimization, and through duality, a set of allocations that lie in the core can be found. Hence, whenever techniques (T1)-(T4) are too complex or difficult to apply for solving a canonical game, as per the *sixth* technique in Table I, one can explore the properties of the considered game model such as in [10], [17–20].

In summary, the core is one of the most important solution concepts in coalitional games, notably in our canonical games class. It must be stressed that the existence of the core shows that the grand coalition $\mathcal{N}$ of a given $(\mathcal{N}, v)$ canonical coalitional game is stable, optimal (from a payoff perspective), and desirable.

## C. The Shapley Value

As a solution concept, the core suffers from three main drawbacks: (i) - The core can be empty, (ii) - the core can be quite large, hence selecting a suitable core allocation can be difficult, and (iii)- in many scenarios, the allocations that lie in the core can be unfair to one or more players. These drawbacks motivated the search for a solution concept which can associate with every coalitional game $(\mathcal{N}, v)$ a *unique* payoff vector known as the *value* of the game (which is quite different from the value of a coalition). Shapley approached this problem axiomatically by defining a set of desirable properties and he characterized a unique mapping $\phi$ that satisfies these axioms, later known as the *Shapley value* [1]. The Shapley value was essentially defined for TU games; however, extensions to NTU games exist. In this tutorial, we restrict our attention to the Shapley value for TU canonical games, and refer the reader to [1, Ch. 9.9] for insights on how the Shapley value is extended to NTU games. Shapley provided four axioms[4] as follows ($\phi_i$ is the payoff given to player $i$ by the Shapley value $\phi$)

1) *Efficiency Axiom*: $\sum_{i \in \mathcal{N}} \phi_i(v) = v(\mathcal{N})$.
2) *Symmetry Axiom*: If player $i$ and player $j$ are such that $v(S \cup \{i\}) = v(S \cup \{j\})$ for every coalition $S$ not containing player $i$ and player $j$, then $\phi_i(v) = \phi_j(v)$.
3) *Dummy Axiom*: If player $i$ is such that $v(S) = v(S \cup \{i\})$ for every coalition $S$ not containing $i$, then $\phi_i(v) = 0$.
4) *Additivity Axiom*: If $u$ and $v$ are characteristic functions, then $\phi(u+v) = \phi(v+u) = \phi(u) + \phi(v)$.

Shapley showed that there exists a unique mapping, the Shapley value $\phi(v)$, from the space of all coalitional games to $\mathbb{R}^N$, that satisfies these axioms. Hence, for every game $(\mathcal{N}, v)$, the Shapley value $\phi$ assigns a unique payoff allocation in $\mathbb{R}^N$ which satisfies the four axioms. The efficiency axiom is in fact group rationality. The symmetry axiom implies that, when two players have the same contribution in a coalition, their assigned payoffs must be equal. The dummy axiom assigns no payoff to players that do not improve the value of any coalition. Finally, the additivity axiom links the value of different games $u$ and $v$ and asserts that $\phi$ is a unique mapping over the space of all coalitional games.

The Shapley value also has an alternative interpretation which takes into account the order in which the players join the grand coalition $\mathcal{N}$. In the event where the players join the grand coalition in a *random* order, the payoff allotted by the Shapley value to a player $i \in \mathcal{N}$ is the expected marginal contribution of player $i$ when it joins the grand coalition.

---

[4]In some references, the Shapley axioms are compressed into three by combining the dummy and efficiency axioms.

The basis of this interpretation is that, given any canonical TU game $(\mathcal{N}, v)$, for every player $i \in \mathcal{N}$ the Shapley value $\phi(v)$ assigns the payoff $\phi_i(v)$ given by

$$\phi_i(v) = \sum_{S \subseteq \mathcal{N} \setminus \{i\}} \frac{|S|!(N - |S| - 1)!}{N!} [v(S \cup \{i\}) - v(S)]. \tag{8}$$

In (8), it is clearly seen that the marginal contribution of every player $i$ in a coalition $S$ is $v(S \cup \{i\}) - v(S)$. The weight that is used in front of $v(S \cup \{i\}) - v(S)$ is the probability that player $i$ faces the coalition $S$ when entering in a random order, i.e., the players in front of $i$ are the ones already in $S$. In this context, there are $|S|!$ ways of positioning the players of $S$ at the start of an ordering, and $(N - |S| - 1)!$ ways of positioning the remaining players except $i$ at the end of an ordering. The probability that such an ordering occurs (when all orderings are equally probable) is therefore $\frac{|S|!(N-|S|-1)!}{N!}$, consequently, the resulting payoff $\phi_i(v)$ is the expected marginal contribution, under random-order joining of the players for forming the grand coalition.

In general, the Shapley value is unrelated to the core. However, in some applications, one can show that the Shapley value lies in the core. Such a result is of interest, since if such an allocation is found, it combines both the stability of the core as well as the axioms and fairness of the Shapley value. In this regard, an interesting result from game theory is that *for convex games the Shapley value lies in the core* [1], [3]. The Shapley value presents an interesting solution concept for canonical games, and has numerous applications in both game theory and communication networks. For instance, in coalitional voting simple games, the Shapley value of a player $i$ represents its power in the game. In such games, the Shapley value is used as a power index (known as the Shapley-Shubik index), and it has a large number of applications in many game theoretical and political settings [3]. In communication networks, the Shapley value presents a suitable fairness criteria for allocating resources or data rates as in [9], [19], [21]. The computation of the Shapley value is generally done using (8); however, in games with a large number of players the computational complexity of the Shapley value grows significantly. For computing the Shapley value in reasonable time, several analytical techniques have been proposed such as multi-linear extensions [3], and sampling methods for simple games [22], among others.

*D. The Nucleolus*

Another prominent and interesting solution concept for canonical games is *the nucleolus* which was introduced mainly for TU games [3]. Extensions of the nucleolus for NTU games are not yet formalized in game theory, and hence this tutorial will only focus on the nucleolus for TU canonical games. The basic motivation behind the nucleolus is that, instead of applying a general fairness axiomatization for finding a unique payoff allocation, i.e., a value for the game, one can provide an allocation that minimizes the dissatisfaction of the players from the allocation they can receive in a given $(\mathcal{N}, v)$ game. For a coalition $S$, the measure of dissatisfaction from an allocation $\boldsymbol{x} \in \mathbb{R}^N$ is defined as the *excess* $e(\boldsymbol{x}, S) = v(S) - \sum_{j \in S} x_j$. Clearly, an allocation $\boldsymbol{x}$ which can ensure that all excesses (or dissatisfactions) are minimized is of particular interest as a solution[5] and hence, constitutes the main motivation behind the concept of the nucleolus. Let $\boldsymbol{O}(\boldsymbol{x})$ be the vector of all excesses in a canonical game $(\mathcal{N}, v)$ arranged in non-increasing order (except the excess of the grand coalition $\mathcal{N}$). A vector $\boldsymbol{y} = (y_1, \ldots, y_k)$ is said to be lexographically less than a vector $\boldsymbol{z} = (z_1, \ldots, z_k)$ (denoted by $\boldsymbol{y} \prec_{\text{lex}} \boldsymbol{z}$) if $\exists l \in \{1, \ldots, k\}$ where $y_1 = z_1, y_2 = z_2, \ldots, y_{l-1} = z_{l-1}, y_l < z_l$. An *imputation* $\boldsymbol{x}$ is a *nucleolus* if for every other imputation $\boldsymbol{\delta}$, $\boldsymbol{O}(\boldsymbol{x}) \prec_{\text{lex}} \boldsymbol{O}(\boldsymbol{\delta})$. Hence, the nucleolus is the imputation $\boldsymbol{x}$ which *minimizes the excesses* in a non-increasing order. The nucleolus of a canonical coalitional game exists and is unique. The nucleolus is group and individually rational (since it is an imputation), and satisfies the symmetry and dummy axioms of Shapley. If the core is not empty, the nucleolus is in the core. Moreover, the nucleolus lies in the *kernel* of the game, which is the set of all allocations $x$ such that $\max_{S\subseteq\mathcal{N}\setminus\{j\},i\in S} e(x,S) = \max_{G\subseteq\mathcal{N}\setminus\{i\},j\in G} e(x,G)$. The kernel states that if players $i$ and $j$ are in the same coalition, then the highest excess that $i$ can make in a coalition without $j$ is equal to the highest excess that $j$ can make in a coalition without $i$. As the nucleolus lies in the kernel, it also verifies this property. Thus, the nucleolus is the best allocation under a min-max criterion. The process for computing the nucleolus is more complex than the Shapley value, and is described as follows. First, we start by finding the imputations that distribute the worth of the grand coalition in such a way that the maximum excess (dissatisfaction) is minimized. In the event where this minimization has a unique solution, this solution is the nucleolus. Otherwise, we search for the imputations which minimize the second largest excess. The procedure is repeated for all subsequent excesses, until finding a unique solution which would be the nucleolus. These sequential minimizations are solved using linear programming techniques such as the simplex method [23]. The applications of the nucleolus are numerous in game theory. One of the most prominent examples is the marriage contract problem which first appeared in the Babylonian Talmud (0-500 A.D).

---

[5]In particular, an imputation **x** lies in the core of $(N, v)$, if and only if all its excesses are negative or zero.

*Example 2: A man has three wives, and he is committed to a marriage contract that specifies that they should receive* 100, 200 *and* 300 *units respectively, after his death. This implies that, given a total amount of* $\alpha$ *units left after the man's death, the three wives can only claim 100, 200, and 300, respectively, out of the* $\alpha$ *units. If after the man dies, the amount of money left is not enough for this distribution, the Talmud recommends the following:*

- *If* $\alpha = 100$ *is available after the man dies, then each wife gets* $\frac{100}{3}$.
- *If* $\alpha = 200$ *is available after the man dies, wife* 1 *gets* 50, *and the other two get* 75 *each*.
- *If* $\alpha = 300$ *is available after the man dies, wife* 1 *gets* 50, *wife* 2 *gets* 100 *and wife* 3 *gets* 150.

*Note that the Talmud does not specify the allocation for other values of* $\alpha$ *but certainly, if* $\alpha \geq 600$ *each wife simply claims its full right. A key question that puzzled mathematicians and researchers in game theory was how this allocation was made and it turns out that the nucleolus is the answer. Let us model the game as a coalitional game* $(\mathcal{N}, v)$ *where* $\mathcal{N}$ *is the set of all three wives which constitute the players and* $v$ *is the value defined for any coalition* $S \subseteq \mathcal{N}$ *as* $v(S) = \max(0, \alpha - \sum_{i\in\mathcal{N}\setminus S} c_i)$, *where* $\alpha \in \{100, 200, 300\}$ *is the total units left after the death of the man and* $c_i$ *is the claim that wife* $i$ *must obtain* ($c_1 = 100, c_2 = 200, c_3 = 300$). *It then turns out that, with this formulation, the payoffs that were recommended by the Talmud coincide with the nucleolus of the game! This result highlights the importance of the nucleolus in allocating fair payoffs in a game.*

In summary, the nucleolus is quite an interesting concept, since it combines a number of fairness criteria with stability. However, the communications applications that utilized the nucleolus are still few, with one example being [19], where it was used for allocating the utilities in the modeled game. The main drawback of the nucleolus is its computational complexity in some games. However, with appropriate models, the nucleolus can be an optimal and fair solution to many applications.

*E. Applications of Canonical Coalitional Games*

*1) Rate allocation in a multiple access channel:* An elegant and interesting use of canonical games within communication networks is presented in [9] for the study of rate allocation in multiple access channels (MAC). The model in [9] tackles the problem of how to fairly allocate the transmission rates between a number of users accessing a wireless Gaussian MAC channel. In this model, the users are bargaining for obtaining a fair allocation of the total transmission rate available. Every user, or group of users (coalition), that does not obtain a fair allocation of the rate can threaten to act on

TABLE II
THE MAIN STEPS IN SOLVING THE GAUSSIAN MAC RATE ALLOCATION CANONICAL GAME AS PER [9]

1- The player set is the set $\mathcal{N}$ of users in a Gaussian MAC channel.
2- For a coalition $S \subseteq \mathcal{N}$, a superadditive value function in characteristic form with TU is defined as the maximum sum-rate (capacity) that $S$ achieves under the assumption that the users in coalition $S^c = \mathcal{N} \setminus S$ attempt to jam the communication of $S$.
3- Through technique (T5) of Table I the core is shown to be non-empty and containing all imputations in the capacity region of the grand coalition.
4- The Shapley value is discussed as a fairness rule for rate-allocation, but is shown to be outside the core, hence, rendering the grand coalition unstable.
5- A new application-specific fairness rule, known as "envy-free" fairness, is shown to lie in the core and is presented as a solution to the rate-allocation game in Gaussian MAC.

its own which can reduce the rate available for the remaining users. Consequently, the game is modeled as a coalitional game defined by $(\mathcal{N}, v)$ where $\mathcal{N} = \{1, \ldots, N\}$ is the set of players, i.e., the wireless network users that need to access the channel, and $v$ is the maximum sum-rate that a coalition $S$ can achieve. In order to have a characteristic function, [9] assumes that, when evaluating the value of a coalition $S \subset \mathcal{N}$, the users in $S^c = \mathcal{N} \setminus S$ known as jammers, cooperate in order to *jam* the transmission of the users in $S$. The jamming assumption is a neat way of maintaining the characteristic form of the game, and it was previously used in game theory for deriving a characteristic function from a strategic form non-cooperative game [1], [12]. Subsequently, when evaluating the sum-rate utility $v(S)$ of any coalition $S \subseteq \mathcal{N}$, the users in $S^c$ form a single coalition to jam the transmission of $S$ and hence, the coalitional structure of $S^c$ is always pre-determined yielding a characteristic form. For a coalition $S$, the characteristic function in [9], $v(S)$, represents the capacity, i.e., the maximum sum-rate, that $S$ achieves under the jamming assumption. Hence, $v(S)$ represents a rate that can be apportioned in an arbitrary manner between the players in $S$, and thus the game is a TU game. It is easily shown in [9] that the game is superadditive since the sum of sum-rates achieved by two disjoint coalitions is no less than the sum-rate achieved by the union of these two coalitions, since the jammer in both cases is the same (due to the assumption of a single coalition of jammers). Consequently, the problem lies in allocating the payoffs, i.e., the transmission rates, between the users in the grand coalition $\mathcal{N}$ which forms in the network. The grand coalition $\mathcal{N}$ has a capacity region $\mathcal{C} = \{\boldsymbol{R} \in \mathbb{R}^N | \sum_{i=1}^N R_i \leq C(\Gamma_S, \sigma^2), \ \forall S \subseteq \mathcal{N}\}$, where $\Gamma_S$ captures the power constraints on the users in $S$, $\sigma^2$ is the Gaussian noise variance, and hence, $C(\Gamma_S, \sigma^2)$ is the maximum sum-rate (capacity) that coalition $S$ can achieve. Based on these properties, the rate allocation game in [9] is clearly a *canonical coalitional game*, and the key question that [9] seeks to answer is "how to allocate the capacity of the grand coalition $v(\mathcal{N})$ among the users in a fair way that stabilizes $\mathcal{N}$". In answering this question, two main concepts from canonical games are used: The core and the Shapley value.

In this rate allocation game, it is shown that the core, which represents the set of rate allocations that stabilize the grand coalition, is *non-empty* using technique (T5) from Table I. By considering the *imputations* that lie in the capacity region $\mathcal{C}$, i.e., the rate vectors $\boldsymbol{R} \in \mathcal{C}$ such that $\sum_{i=1}^N R_i = C(\Gamma_\mathcal{N}, \sigma^2)$, it is shown that any such vector lies in the core. Therefore, the grand coalition $\mathcal{N}$ of the Gaussian MAC canonical game can be stabilized. However, the core of this game is *big* and contains a large number of rate vectors. Thus, the authors in [9] sought to answer the next question "how to select a single fair allocation which lies in the core?". For this purpose, the authors investigate the use of the Shapley value as a fair solution for rate allocation which accounts for the random-order of joining of the players in the grand coalition. In this setting, the Shapley value simply implies that no rate is left unallocated (efficiency axiom), dummy players receive no rate (dummy axiom), and the labeling of the players does not affect the rate that they receive (symmetry axiom). However, the authors show that: (i)- The fourth Shapley axiom (additivity) is not suitable for the proposed rate allocation game, and (ii)- the Shapley value does not lie in the core, and hence cannot stabilize the grand coalition. Based on these results for the Shapley value, the authors propose a new fairness criterion, named "envy-free" fairness. The envy-free fairness criterion relies on the first three axioms of Shapley (without the additivity axiom), and complements them with a fourth axiom, the *envy free allocation axiom* [9, Eq. (6)]. This axiom states that, given two players $i$ and $j$, with power constraints $\Gamma_i > \Gamma_j$, an envy-free allocation $\psi$ gives a payoff $\psi_j(v)$ for user $j$ in the game $(\mathcal{N}, v)$, equal to the payoff $\psi_i(v^{i,j})$ of user $i$ in the game $(\mathcal{N}, v^{i,j})$ where $v^{i,j}$ is the value of the game where user $i$ utilizes a power $\Gamma_i = \Gamma_j$. Mathematically, this axiom implies that $\psi_j(v) = \psi_i(v^{i,j})$. With these axioms, it is shown that a unique allocation exists and this allocation lies in the core. Thus, the envy-free allocation is presented as a fair and suitable solution for the rate allocation game in [9]. Finally, the approach used for solving the rate allocation canonical coalitional game in [9] is summarized in Table II.

*2) Canonical games for receivers and transmitters cooperation:* In [10], canonical games are used for studying the cooperation possibilities between single antenna receivers and transmitters in an interference channel. The model considered in [10] consists of a set of transmitter-receiver pairs, in a Gaussian interference channel. The authors study the cooperation between the receivers under two coalitional game models: A TU model where the receivers communicate through noise-free channels and jointly decode the received signals, and an NTU model where the receivers cooperate by forming a linear multiuser detector (in this case the interference channel is reduced to a MAC channel). Further, the authors study the transmitters cooperation problem under perfect cooperation and partial decode and forward cooperation, while considering that the receivers have formed the grand coalition. Since all the considered games are canonical (as we will see later), the main interest is in studying the properties of the grand coalitions for the receivers and the transmitters.

For receiver cooperation using joint decoding, the coalitional game model is as follows: the player set $\mathcal{N}$ is the set of links (the players are the receivers of these links) and, assuming that the transmitters do *not* cooperate, the value $v(S)$ of a coalition $S \subseteq \mathcal{N}$ is the maximum sum-rate achieved by the links whose receivers belong to $S$. Under this model, one can easily see that the utility is transferable since it represents a sum-rate, hence the game is TU. The game is also in characteristic form, since, as the transmitters are considered non-cooperative, the sum-rate achieved when the receivers in $S$ cooperate depends solely on the receivers in $S$ while treating the signal from the links in $\mathcal{N} \setminus S$ as interference. In this game, the cooperation channels between the receivers are considered noiseless and hence, cooperation is always beneficial and the game is shown to be superadditive. Hence, under our proposed classification, this game is clearly a canonical game, and the interest is in studying the properties of the grand coalition of receivers. Under this cooperation scheme, the network can be seen as a single-input-multiple-output (SIMO) MAC channel, and the proposed coalitional game is shown to have a non-empty core which contains *all the imputations* which lie on the SIMO-MAC capacity region. The technique used for this proof is similar to the game in [9] which selects a particular set of rate vectors, those that are on the SIMO-MAC region, and shows that they lie in the core as per (T5) from Table I. The core of this game is very large, and for selecting fair allocations, it is proven in [10] that the Nash bargaining solution, and in particular, a proportional fair rate allocation lie in the core, and hence constitute suitable fair and stable allocations. For the second receiver cooperation game, the model is similar to the joint decoding game, with one major difference: Instead of jointly decoding the received signals, the receivers form linear multiuser detectors (MUD). The MUD coalitional game is inherently different from the joint decoding game since, in a MUD, the SINR ratio achieved by a user $i$ in coalition $S$ cannot be shared with the other users, and hence the game becomes an NTU game with the SINR representing the payoff of each player. In this NTU setting, the value $v(S)$ of a coalition $S$ becomes the set of SINR vectors that a coalition $S$ can achieve. For this NTU game, the grand coalition is proven to be stable and sum-rate maximizing at high SINR regime using limiting conditions on the SINR expression, hence technique (T6) in Table I.

TABLE III
THE MAIN RESULTS FOR RECEIVERS AND TRANSMITTERS COOPERATION COALITIONAL GAMES AS PER [10]

| |
|---|
| 1- The coalitional game between the receivers, where cooperation entails joint decoding of the received signal, is a canonical TU game which has a non-empty core. Hence, the grand coalition is the stable and sum-rate maximizing coalition. |
| 2- The coalitional game between the receivers, where cooperation entails forming linear multiuser detectors, is a canonical NTU game which has a non-empty core. Hence, the grand coalition is the stable and sum-rate maximizing coalition. |
| 3- For transmitters cooperation, under jamming assumption, the coalitional game is not *superadditive*, hence non-canonical. However, the grand coalition is shown to be the rate maximizing partition. |
| 4- For transmitters cooperation under jamming assumption, no results for the existence of the core can be found due to mathematical intractability. |

For modeling the transmitters cooperation problem as a coalitional game the authors make two assumptions: (i)- The receivers jointly decode the signals, hence form a grand coalition, and (ii)- a jamming assumption similar to [9] is made for the purpose of maintaining the characteristic form. In the transmitters game, from the set of links $\mathcal{N}$, the transmitters are the players. When considering the transmitters cooperation along with the receivers cooperation, the interference channel is mapped unto a multiple-input-multiple-output (MIMO) MAC channel. For maintaining a characteristic form, the authors assumed, in a manner analogous to [9], that whenever a coalition of transmitters $S$ forms, the users in $S^c = \mathcal{N} \setminus S$ form one coalition and aim to jam the transmission of coalition $S$. Without this assumption, the maximum sum-rate that a coalition can obtain highly depends on how the users in $S^c$ structure themselves, and hence requires a partition form that may be difficult to solve. With these assumptions, the value of a coalition $S$ is defined as the maximum sum-rate achieved by $S$ when the coalition $S^c$ seeks to jam the transmission of $S$. Using this transmitters with jamming coalitional game, the authors show that in general the game has an empty core. This game is not totally canonical since it does not satisfy the superadditivity property. However, by proving through [10, Th. 19] that the grand coalition is the optimal partition, from a total utility point of view, the grand coalition becomes the main candidate partition for the core. The authors conjecture that in some cases, the core can also be non-empty depending on the power and channel gains. However, no existence results for the core are provided in this game. Finally, the authors in [10] provide a discussion on the grand coalition and its feasibility when the transmitters employ a partial decode and forward cooperation. The main results are summarized in Table III.

*3) Other applications for canonical games and future directions:* Canonical coalitional games cover a broad range of communication and networking applications and, indeed, most research activities in these areas utilize the tools that fall under the canonical coalitional games class. In addition to the previous examples, numerous applications used models that involve canonical games. For instance, in [19], canonical coalitional games are used to solve an inherent problem in packet forwarding ad hoc networks. In such networks, the users that are located in the center of the network, known as backbone nodes, have a mutual benefit to forward each others' packets. In contrast, users located at the boundary of the network, known as boundary nodes, are not helped by the backbone nodes due to the fact that the backbone nodes do not need the help of the boundary nodes at any time. Hence, in such a setting, the boundary nodes end up having no way of sending their packets to other nodes, and this is a problem known as the *curse of the boundary nodes*. In [19], a canonical coalitional game model is proposed between a player set $\mathcal{N}$ which includes all boundary nodes and a *single* backbone node. In this model, forming a coalition, entails the following benefits: (i)- By cooperating with a number of boundary nodes and using cooperative transmission, the backbone node can reduce its power consumption, and (ii)- in return, the backbone node agrees to forward the packets of the boundary nodes. For cooperative transmission, in a coalition $S$, the boundary nodes act as relays while the backbone node acts as a source. In this game, the *core* is shown to be non-empty using the property that any group of boundary nodes receive no utility if they break away from the grand coalition with the backbone node, this classifies as a (T6) technique from Table I. Further, the authors in [19] study the conditions under which a Shapley value and a nucleolus are suitable for modeling the game. By using a canonical game, the connectivity of the ad hoc network is significantly improved [19]. Beyond packet forwarding, many other applications such as in [17], [18], [21] utilize several of the techniques in Table I for studying the grand coalition in a variety of communications applications.

In summary, canonical games are an important tool for studying cooperation and fairness in communication networks, notably when cooperation is always beneficial. Future applications are numerous, such as studying cooperative transmission capacity gains, distributed cooperative source coding, cooperative relaying in cognitive radio and many other applications. In brief, whenever a cooperative scheme that yields significant gains at any layer is devised, one can utilize canonical coalitional games for assessing the stability of the grand coalition and identifying fairness criteria in allocating the gains that result from cooperation. Finally, it has to be noted that canonical games are not restricted to link-level analysis, but also extend to network-level studies as demonstrated in [18], [19].

## IV. CLASS II: COALITION FORMATION GAMES

### A. Main Properties of Coalition Formation Games

Coalition formation games encompass coalitional games where, unlike the canonical class, *network structure* and *cost* for cooperation play a major role. Some of the main characteristics that make a game a coalition formation game are as follows:

1) The game is in either characteristic form or partition form (TU or NTU), and is generally not superadditive.
2) Forming a coalition brings gains to its members, but the gains are limited by a *cost* for forming the coalition, hence the grand coalition is seldom the optimal structure.
3) The objective is to study the *network coalitional structure*, i.e., answering questions like which coalitions will form, what is the optimal coalition size and how can we assess the structure's characteristics, and so on.
4) The coalitional game is subject to environmental changes such as a variation in the number of players, a change in the strength of each player or other factors which can affect the network's topology.
5) A coalitional structure is imposed by an external factor on the game (e.g., physical restrictions in the problem).

Unlike canonical games, a coalition formation game is generally not superadditive and can support the partition form model. Another important characteristic which classifies a game as a coalition formation game is the presence of a cost for forming coalitions. In canonical games, as well as in most of the literature, there is an implicit assumption that forming a coalition is always beneficial (e.g. through superadditivity). In many problems, forming a coalition requires a negotiation process or an information exchange process which can incur a cost, thus, reducing the gains from forming the coalition. In general, coalition formation games are of two types: *Static coalition formation games and dynamic coalition formation games*. In the former, an external factor imposes a certain coalitional structure, and the objective is to study this structure. The latter is a more rich framework. In dynamic coalition formation games, the main objectives are to analyze the formation of a coalitional structure, through players' interaction, as well as to study the properties of this structure and its adaptability to environmental variations or externalities. In contrast to canonical games, where formal rules and analytical concepts exist, solving a coalition formation game, notably dynamic coalition formation, is more difficult, and application-specific. The rest of this section is devoted to dissecting the key properties of coalition formation games.

*B. Impact of a Coalitional Structure on Solution Concepts of Canonical Coalitional Games*

In canonical games, the solution concepts defined, such as the core, the Shapley value and the nucleolus, assumed that the grand coalition would form due to the superadditivity property. The presence of a coalitional structure affects the definition and use of these concepts as was first pointed out by Aumann and Drèze in [24] for a static coalition formation game. In [24], a TU coalitional game is considered, in the presence of a static coalitional structure $\mathcal{B} = \{B_1, \ldots, B_l\}$ (each $B_i$ is a coalition), that is imposed by some external factor. Hence, [24] defines a coalitional game as the triplet $(\mathcal{N}, v, \mathcal{B})$ where $v$ is a characteristic function. First, in the presence of $\mathcal{B}$, the concept of group rationality is substituted by *relative efficiency*. Given an allocation vector $\boldsymbol{x} \in \mathbb{R}^N$, relative efficiency implies that, for each coalition $B_k \in \mathcal{B}$, $\sum_{i \in B_k} x_i = v(B_k)$ [24]. Hence, for every present coalition $B_k$ in $\mathcal{B}$, the total value available for coalition $B_k$ is divided among its members unlike in canonical games where the value of the grand coalition $v(\mathcal{N})$ is distributed among all players. With regards to canonical solutions, we first turn our attention to the Shapley value. For the game $(\mathcal{N}, v, \mathcal{B})$, the previously defined Shapley axioms remain in place, except for the efficiency axiom which is replaced by a *relative efficiency* axiom. With this modified axiom, [24] shows that the Shapley value of $(\mathcal{N}, v, \mathcal{B})$, referred to as $\mathcal{B}$-value, has the *restriction property*. The restriction property implies that, for finding the $\mathcal{B}$-value, one can consider the *restricted coalitional games* $(B_k, v|B_k), \forall B_k \in \mathcal{B}$ where $(v|B_k)$ is the value $v$ of the original game $(\mathcal{N}, v, \mathcal{B})$, defined over player set (coalition) $B_k$. As a result, for finding the $\mathcal{B}$-value, we proceed in two steps, using the restriction property: (1)- Consider the games $(B_k, v|B_k), k = 1, \ldots, l$ separately and for each such game $(B_k, v|B_k)$ find the Shapley value using the canonical definition (8), and (2)- the $\mathcal{B}$-value of the game is the $1 \times N$ vector $\boldsymbol{\phi}$ of payoffs constructed by combining the resulting allocations of each restricted game $(B_k, v|B_k)$.

In the presence of a coalitional structure $\mathcal{B}$, the canonical definitions of the core and the nucleolus are also mainly modified by replacing group rationality with relative efficiency. However, unlike the Shapley value, it is shown in [24] that the restriction property does not apply to the core, nor the nucleolus. This can be easily deduced from the fact that both the core and the nucleolus depend on *all* coalitions of $\mathcal{N}$. Hence, in the presence of $\mathcal{B}$, the core and the nucleolus depend on the values of coalitions $B_j \in \mathcal{B}$ *as well as* the values of coalitions that are not in $\mathcal{B}$, that is coalitions $S \subset \mathcal{N}, \nexists B_k \in \mathcal{B}$ s. t. $B_k = S$. Hence, the problem of finding the core and the nucleolus of $(\mathcal{N}, v, \mathcal{B})$ is more complex than for the Shapley value. In [24], an approach for finding these solutions for games where $v(\{i\}) = 0, \forall i \in \mathcal{N}$ is presented. The approach is based on finding a game equivalent to $v$ by redefining the value, and hence, the core and nucleolus can be found for this equivalent game. For the detailed analysis, we refer the reader to [24, Th. 4 and Th. 5].

Even though the analysis in [24] is restricted to static coalition formation games with TU and in characteristic form, it shows that finding solutions for coalition formation games is not straightforward. The difficulty of such solutions increases whenever an NTU game, a partition form game, or a dynamic coalition formation game are considered, notably when the objective is to compute the solution in a distributed manner. For example, when considering a dynamic coalition formation game, one would need to evaluate the payoff allocations *jointly* with the formation of the coalitional structure, hence solution concepts become even more complex to find (although the restriction property of the Shapley value makes things easier). For this purpose, the literature dealing with coalition formation games, notably dynamic coalition formation such as [25–28] or others, usually re-defines the solution concepts or presents alternatives that are specific to the game being studied. Hence, unlike canonical games where formal solutions exist, the solution of a coalition formation game depends on the model and the objectives that are being considered.

*C. Dynamic Coalition Formation Algorithms*

In general, in a coalition formation game, the most important aspect is the formation of the coalitions in the game. In other words, one must answer the question of "how to form a coalitional structure that is suitable to the studied game". In addition, the evolution of this structure is important, notably when changes to the game nature can occur due to external or internal factors (e.g., what happens to the coalition structure if one or more players leave the game). In many applications, coalition formation entails finding a coalitional structure which either maximizes the total utility (social welfare) if the game is TU, or finding a structure with Pareto optimal payoff distribution for the players if the game is NTU. For achieving such a goal, a *centralized* approach can be used; however, such an approach is generally NP-complete [25–28]. The reason is that, finding an optimal partition, requires iterating over all the partitions of the player set $\mathcal{N}$. The number of partitions of a set $\mathcal{N}$ grows exponentially with the number of players in $\mathcal{N}$ and is given by a value known as the Bell number [25]. For example, for a game where $\mathcal{N}$ has only 10 elements, the number of partitions that a centralized approach must go through is 115975 (easily computed through the Bell number). Hence, finding an optimal partition from a centralized approach is, in general, computationally complex and impractical. In some cases, it may be possible to explore the properties of the game, notably of the value $v$, for reducing the centralized complexity. Nonetheless, in many practical applications, it is desirable that the coalition formation process takes place in a distributed manner, whereby the players have an autonomy on the decision as to whether or not they join a coalition. In fact, the complexity of the centralized approach as well as the need for distributed solutions have sparked a huge growth in the coalition formation literature that aims to find low complexity and distributed algorithms for forming coalitions [25–28].

The approaches used for distributed coalition formation are quite varied and range from heuristic approaches [25], Markov chain-based methods [26], to set theory based methods [27] as well as approaches that use bargaining theory or other negotiation techniques from economics [28]. Although there are no general rules for distributed coalition formation, some work, such as [27] provides generic rules that can be used to derive application-specific coalition formation algorithms. Although [27] does not explicitly construct a coalition formation algorithm, the mathematical framework presented can be used to develop such algorithms. The main ingredients that are presented in [27] are three: (1)- Well-defined orders suitable to compare *collections* of coalitions, (2)- two simple rules for forming or breaking coalitions, and (3)- adequate notions for assessing the stability of a partition. For comparing collections of coalitions, a number of *orders* are defined in [27], two of which are of noticeable importance. The first order, known as the *utilitarian order*, states that, a group of players prefers to organize themselves into a collection $\mathcal{R} = \{R_1, \ldots, R_k\}$ instead of $\mathcal{S} = \{S_1, \ldots, S_l\}$, if the total social welfare achieved in $\mathcal{R}$ is strictly greater than in $\mathcal{S}$, i.e., $\sum_{i=1}^{k} v(R_i) > \sum_{i=1}^{l} v(S_i)$. This order is generally suitable for TU games. Another important order is the *Pareto order*, which bases the preference on the individual payoffs of the players rather than the coalition value. Given two allocations $\boldsymbol{x}$ and $\boldsymbol{y}$ that are allotted by $\mathcal{R}$ and $\mathcal{S}$, respectively, to the same players, $\mathcal{R}$ is preferred over $\mathcal{S}$ by Pareto order if at least one player improves in $\mathcal{R}$ without hurting the other players, i.e., $\boldsymbol{x} \geq \boldsymbol{y}$ with at least one element $x_i$ of $\boldsymbol{x}$ such that $x_i > y_i$. The Pareto order is suitable for both TU and NTU games.

Using such orders, [27] presents two main rules for forming or breaking coalitions, referred to as *merge* and *split*. The basic idea behind the rules is that, given a set of players $\mathcal{N}$, any collection of disjoint coalitions $\{S_1, \ldots, S_l\}$, $S_i \subset \mathcal{N}$ can agree to *merge* into a single coalition $G = \cup_{i=1}^{l} S_i$, if this new coalition $G$ is preferred by the players over the previous state depending on the selected comparison order. Similarly, a coalition $S$ *splits* into smaller coalitions if the resulting collection $\{S_1, \ldots, S_l\}$ is preferred by the players over $S$. Independent of the selected order, any arbitrary sequence of these two rules is shown to converge to a final partition of $\mathcal{N}$ [27]. For assessing the stability of the final partition, the authors in [27] propose the concept of a *defection function*, which is a function that associates with every network partition, another partition, a group of other partitions, or a group of collections in $\mathcal{N}$. By defining various types of such a function, one can assess whether, in a given partition $\mathcal{T}$ of $\mathcal{N}$, there is an incentive for the players to deviate and form other partitions or collections. A first notion of stability, is a weak equilibrium-like stability, known as $\mathbb{D}_{hp}$ stability. A $\mathbb{D}_{hp}$-stable partition simply implies that, in this partition, no group of players has an interest in performing a merge or a split operation. This type of stability can be thought of as merge-and-split proofness of a partition, or a kind of equilibrium with respect to merge-and-split. The most important type of stability inspected in [27] is $\mathbb{D}_c$-stability. The existence of a $\mathbb{D}_c$-stable partition is not always guaranteed, and the two conditions needed for its existence can be found in [27]. However, when it exists, the $\mathbb{D}_c$-stable partition has numerous attractive properties. First and foremost, a $\mathbb{D}_c$-stable partition is a *unique* outcome of any arbitrary merge and split iteration. Hence, starting from any given partition, one would always reach the $\mathbb{D}_c$-stable partition by merge-and-split. Based on the selected order, the players prefer the $\mathbb{D}_c$-stable partition over *all other partitions*. On one hand, if the selected order is the utilitarian order, this implies that the $\mathbb{D}_c$-stable partition maximizes the social welfare (total utility), on the other hand, if the selected order is the Pareto order, the $\mathbb{D}_c$-stable partition has a Pareto optimal payoff distribution for the players. Finally, no group of players in a $\mathbb{D}_c$-stable partition have an incentive to leave this partition for forming any other collection in $\mathcal{N}$. Depending on the application being investigated, one can possibly define other suitable defection functions, as this concept is not limited to a particular problem.

Coalition formation games are diverse, and by no means limited to the concepts in [27]. For example, a type of coalition formation games, known as *hedonic coalition formation* games has been widely studied in game theory. Hedonic games are quite interesting since they allow the formation of coalitions (whether dynamic or static) based on the individual preferences of the players. In addition, these games admit different stability concepts that are extensions to well known concepts such as the core or the Nash equilibrium used in a coalition formation setting [29]. In this regard, hedonic games constitute a very useful analytical framework which has a very strong potential to be adopted in modeling problems in wireless and communication networks (only few contributions such as [30] used this framework in a communication/wireless model). Furthermore, beyond merge-and-split and hedonic games, dynamic coalition formation games encompass a multitude of algorithms and concepts such as in [25–28] and many others. Due to space limitations, this tutorial cannot provide an exhaustive survey of all such algorithms. Nonetheless, as will be seen in the following sections, many coalition formation algorithms and concepts can be tailored and adapted for communication applications.

### D. Applications of Coalition Formation Games

*1) Transmitter cooperation with cost in a TDMA system:* The formation of virtual MIMO systems through distributed cooperation has received an increasing attention recently (see [10], [31] and the references therein). The problem involves a number of single antenna users which cooperate and share their antennas in order to benefit from spatial diversity or multiplexing, and hence form a virtual MIMO system. Most literature that studied the problem is either devoted to analyzing the link-level information theoretical gains from distributed cooperation, or focused on assessing the stability of the grand coalition, for cooperation with no cost, such as in the work of [10] previously described. However, there is a lack of literature which studies how a network of users can interact to form virtual MIMO systems, notably when there is a cost for cooperation. Hence, a study of the network topology and dynamics that result from the interaction of the users is needed and, for this purpose, coalition formation games are quite an appealing tool. These considerations motivated our work in [31] where we considered a network of single antenna transmitters that send data in the uplink of a TDMA system to a receiver with multiple antennas. In a non-cooperative approach, each single antenna transmitter sends its data in an allotted slot. For improving their capacity, the transmitters can interact for forming coalitions, whereby each coalition $S$ is seen as a single user MIMO that transmits in the slots that were previously held by the users of $S$. After cooperation, the TDMA system schedules one coalition per time slot. An illustration of the model is shown in Fig. 3. To cooperate, the transmitters must exchange their data, and hence, this exchange of information incurs a cost in terms of power. The presence of this cost, as per [31], renders the game non-superadditive due to the fact that the information exchange incurs a cost in power which is increasing with the distances inside the coalition as well as the coalition size. For example, when two users are far away, information exchange can consume the total power, and the utility for cooperation is smaller than in the non-cooperative case. Similarly, adding more users to a coalition does not always yield an increase in the utility; for instance, a coalition consisting of a large number of users increases the cost for information exchange, and thus superadditivity can not be guaranteed. As a consequence of this property, for the proposed game in [31] the grand coalition seldom forms[6] and the game is modeled as a dynamic coalition formation game between the transmitters (identified by the set $\mathcal{N}$) that seek to form cooperating coalitions. The dynamic aspect stems from the fact that many environmental changes, such as the mobility of the transmitters or the deployment of new users, may affect the coalitional structure that will form and any algorithm must be able to cope with these changes accurately.

For the proposed game, the value function represents the sum-rate, or capacity, that the coalition can achieve, while taking into account the power cost. Due to the TDMA nature of the problem, a power constraint $\tilde{P}$ per time slot, and hence per coalition, is considered. Whenever a coalition forms, a fraction of $\tilde{P}$ is used for information exchange, hence constituting a cost for cooperation, while the remaining fraction will be used for the coalition to transmit its data, as a single user MIMO, to the receiver. For a coalition $S$, the fraction used for information exchange is the sum of the powers that each user $i \in S$ needs to transmit its data to the user $j \in S$ that is farthest from $i$; due to the broadcast nature of the wireless channel all other users in $S$ can receive this data as well. This power cost scales with the number of users in the coalition, as well as the distance between these users. Hence, the sum-rate that a coalition can achieve is limited by the fraction of power spent for information exchange. For instance, if the power for information exchange for a coalition $S$ is larger than $\tilde{P}$, then $v(S) = 0$. Otherwise, $v(S)$ represents the sum-rate achieved by the coalition using the remaining fraction of power. Clearly, the sum-rate is a transferable utility, and hence we deal with a TU game.

In this framework, a dynamic coalition formation algorithm based on the merge-and-split rules previously described can be built. In [31], for coalition formation, we start with a non-cooperative network, whereby each user discovers its neighbors starting with the closest, and attempts to merge based on the utilitarian order, i.e., if cooperating with a neighbor improves the total sum-rate that the involved users can achieve, then merging occurs (merge is done through pairwise interactions between a user or coalition and the users or coalitions in the vicinity). Further, if a formed coalition finds out that splitting into smaller coalitions improves the total utility achieved by its users, then a split occurs. Starting from the initial non-cooperative network, the coalition formation algorithm involves sequential merge and split rules. The network's coalition can autonomously decide on whether to perform a merge or split based on the utility evaluation. The convergence is guaranteed by virtue of the definition of merge-and-split. Further, if an optimal $\mathbb{D}_c$-stable partition exists, the proposed algorithm converges to it. The existence of the $\mathbb{D}_c$-stable partition in this model cannot always be guaranteed, as it depends on random locations of the users; however, the convergence to it, when it exists, is guaranteed. The coalition formation algorithm proposed in [31] can handle any network size, as the implementation is inherently distributed, whereby each coalition (or user) can detect the strength of the other users' uplink signals (using techniques as in ad hoc routing), and discover the nearby candidate partners. Consequently, the distributed users can exchange the required information and then assess what kind of merge or split decisions they can make. The transmitters engage in merge-and-split periodically, and hence, adapt the topology to any environmental change, such as mobility or the joining/leaving of transmitters. In this regard, by adequate merge or split decisions, the topology is always dynamically changing, through individual and distributed decisions by the network's coalitions. As the proposed model is TU, several rules for dividing the coalition's value are used. These rules range from well-known fairness criteria such as the proportional fair division, to coalitional game-specific rules such as the Shapley value or the nucleolus. Due to the distributed nature of the problem, the nucleolus or the Shapley value are applied at the level of the coalitions that are forming or splitting. Hence, although for the Shapley value this allocation coincides with the Shapley value of the whole game as previously discussed, for the nucleolus, the resulting allocations lie in the nucleolus of the restricted games only. In this game, for any coalition $S \subseteq \mathcal{N}$ that forms through merge-and-split, the Shapley value presents a division of the payoff that takes into account the random order of joining of the transmitter in $S$ when forming the coalition (this division is also efficient at the coalition level and treats the players symmetrically within $S$). In contrast, the division by the nucleolus at the level of every coalition $S \subseteq \mathcal{N}$ that forms through merge-and-split ensures that the dissatisfaction of any transmitter within $S$ is minimized by minimizing the excesses inside $S$. Finally, although in [31] we used a utilitarian order, in extensions to the work, we reverted to the Pareto order, which allows every user of the coalition to assess the improvement to its own payoff during merge or split, instead of relying on the entire coalitional value. By doing so, the fairness criteria chosen impacts the network structure and hence, for every fairness type one can obtain a different topology.

---

[6]In this game, the grand coalition only forms in extremely favorable cases, such as when the network contains only two users and these users are very closeby.

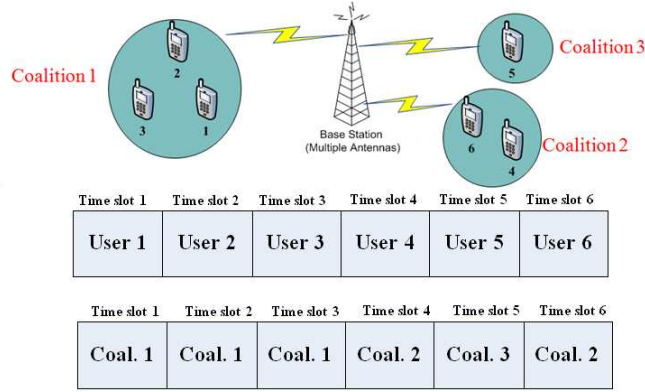

Fig. 3. The system model for the virtual MIMO formation game in [31].

*2) Coalition formation for spectrum sensing in cognitive radio networks:* In cognitive radio networks, the unlicensed secondary users (SU) are required to sense the environment in order to detect the presence of the licensed primary user (PU) and transmit during periods where the PU is inactive. Collaborative spectrum sensing (CSS) has been proposed for improving the sensing performance of the SUs, in terms of reducing the probability of missing the detection of the PU (probability of miss), and hence decreasing the interference on the PU. Even though CSS decreases the probability of miss, it also increases the false alarm probability, i.e., the probability of falsely detecting that the PU is transmitting. Hence, CSS presents an inherent tradeoff between reducing the probability of miss (reducing interference on the PU) and maintaining a good false alarm probability, which corresponds to a good spectrum utilization. In [32], we consider a network of SUs, that interact for improving their sensing performance, while taking into account the false alarm cost. For performing CSS, every group of SUs form a coalition, and within each coalition, an SU, selected as coalition head will gather the sensing bit from the coalition members. By using well-known decision fusion rules, the coalition head can decide on the presence or the absence of the PU. Using this CSS scheme, as shown in [32], each coalition reduces the probability of miss of its SUs. However, this reduction is accompanied by an increase in the false alarm probability. This tradeoff between the improvement of the probability of miss and the false alarm, impacts the coalitional structure that forms in the network.

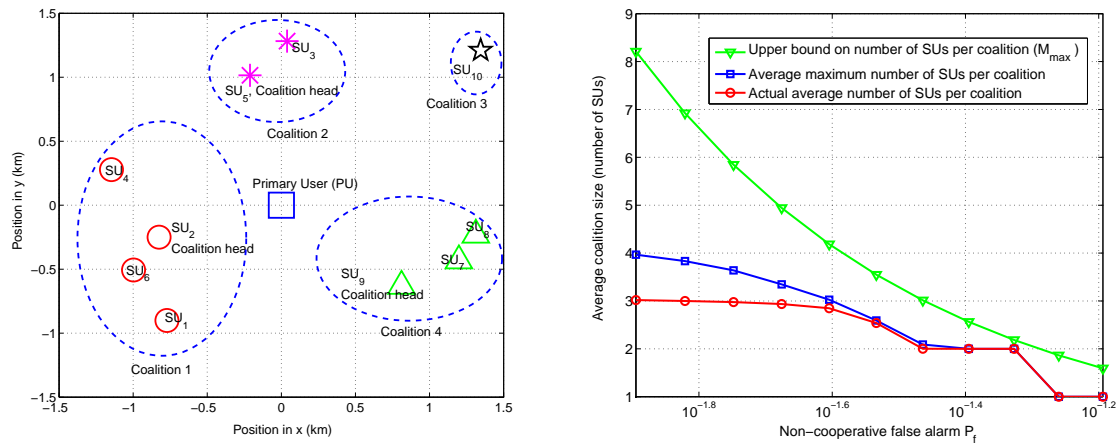

Fig. 4. (a) Topology resulting from coalition formation in CSS for 10 SUs. (b) Maximum and average coalition size vs. non-cooperative false alarm $P_f$ for the dynamic coalition formation game solution for a network of 30 SUs.

Consequently, the CSS problem is modeled as a dynamic coalition formation game between the SUs ($\mathcal{N}$ is the set of SUs in this game). The utility $v(S)$ of each coalition $S$ is a decreasing function of the probability of miss $Q_{m,S}$ within coalition $S$ and a decreasing function of the false alarm probability $Q_{f,S}$. In the false alarm cost component, the proposed utility in [32, Eq. (8)] imposes a maximum tolerable false alarm probability, i.e., an upper bound constraint $\alpha$ on the false alarm, that cannot be exceeded by any SU. This utility represents probabilities, and hence, cannot be transferred arbitrarily between the SUs. Hence, the coalition formation game for CSS is an NTU game, whereby the payoff of an SU which is a member of any coalition $S$ is given by $x_i = v(S), \ \forall i \in S$ and reflects the probabilities of miss and false alarm that any SU which is a member of $S$ achieves [32, Property 1] (here, the NTU value is a singleton set). In this game, it is easily shown that the grand coalition *seldom* forms, due to the false alarm constraint $\alpha$ and the fact that the false alarm for a coalition increases with the coalition size and the distances between the coalition members [32, Property 3].

For this purpose, a coalition formation algorithm is needed. The algorithm proposed in [32] consists of three phases: In the first phase the SUs perform their local sensing, in the second phase the SUs engage in an adaptive coalition formation algorithm based on the merge and split rules of Section IV-C, and in the third phase, once the coalitions have formed, each SU reports its sensing bit to the coalition head which makes a decision on whether or not the PU is present. Due to the NTU nature of the game, the adaptive coalition formation phase of the algorithm uses the Pareto order for performing merge or split operations. The merge and split decisions in the context of the CSS model can also be performed in a distributed manner by each coalition, or individual SU. The merge and split phase converges to the $\mathbb{D}_c$-stable partition which leads to a Pareto optimal payoff allocation, whenever this partition exists. Periodically, the formed coalitions engage in merge and split operations for adapting the topology to environmental changes such as the mobility of the SUs or the PU, or the deployment of more SUs. In Fig. 4 (a), we show an example of a coalitional structure that the SUs form for CSS in a cognitive network of 10 SUs with a false alarm constraint of $\alpha = 0.1$. Clearly, the proposed algorithm allows the SUs to structure themselves into disjoint independent coalitions for the purpose of spectrum sensing. By forming such topologies, it is shown in [32] that the SUs can significantly improve their performance, in terms of probability of miss, reaching up to $86.6\%$ per SU improvement relative to the non-cooperative sensing case for a network of 30 SUs, while maintaining the desired false alarm level of $\alpha = 0.1$. In addition to the performance improvement achieved by the proposed coalition formation algorithm in [32], an interesting upper bound on the coalition size is derived for the proposed utility. This upper bound is a function of only two quantities: The false alarm constraint $\alpha$ and the non-cooperative false alarm value $P_f$, i.e., the detection threshold. Hence, this upper bound does not depend on the location of the SUs in the network nor on the actual number of SUs in the network. Therefore, deploying more SUs or moving the SUs in the network for a fixed $\alpha$ and $P_f$ does not increase the upper bound on coalition size. In Fig. 4 (b), we show this upper bound in addition to the average and maximum achieved coalition size for a network of 30 SUs with a false alarm constraint of $\alpha = 0.1$. The coalition size variations are shown as a function of the non-cooperative false alarm $P_f$. The results in Fig. 4 (b) show that, in general, the network topology is composed of a large number of small coalitions rather than a small number of large coalitions, even when $P_f$ is small relative to $\alpha$ and the collaboration possibilities are high (a smaller $P_f$ implies the cost for cooperation, in terms of false alarm increases more slowly with the coalition size). Also, when $P_f = \alpha = 0.1$, the network is non-cooperative, since cooperation would always violate the false alarm constraint $\alpha$. In a nutshell, dynamic coalition formation provides novel collaboration strategies for SUs in a cognitive network which are seeking to improve their sensing performance, while maintaining a desired spectrum utilization (false alarm level). The framework of dynamic coalition formation games suitably models this problem, yields a significant performance improvement, and allows to characterize the network topology that will form.

*3) Future applications of coalition formation games:* Potential applications of coalition formation games in communication networks are numerous and diverse. Beyond the applications presented above, coalition formation games have already been applied in [33] to improve the physical layer security of wireless nodes through cooperation among the transmitters, while in [30] coalition formation among a number of autonomous agents, such as unmanned aerial vehicles, is studied in the context of data collection and transmission in wireless networks. Moreover, recently, there has been a significant increase of interest in designing autonomic communication systems. Autonomic systems are networks that are self-configuring, self-organizing, self-optimizing, and self-protecting. In such networks, the users should be able to learn and adapt to their environment (changes in topology, technologies, service demands, application context, etc), thus providing much needed flexibility and functional scalability. Coalition formation games present an adequate framework for the modeling and analysis of these self-organizing next generation communication networks. Hence, potential applications of coalition formation games encompass cooperative networks, wireless sensor networks, next generation IP networks, ad hoc self-configuring networks, and many others. In general, whenever there is a need for distributed algorithms for autonomic networks, coalition formation is a strong tool for modeling such problems. Also, any problem involving the study of cooperative wireless nodes behavior when a cost is present, is candidate for modeling using coalition formation games.

Finally, although the main applications we described in this tutorial required a characteristic form, coalition formation games in partition form are of major interest and can have potential applications in communication networks. For instance, in [10], the transmitter cooperation problem assumed that the players outside any coalition work as a single entity and jam the communication of this coalition. This assumption is made in order to have a characteristic form. For relaxing this assumption and taking into account the actual interference that affects a coalition, a coalitional game in partition form is needed. In the presence of a cooperative cost, this partition form game falls in the class of coalition formation games. Hence, coalition formation games in partition form are ripe for many future applications.

## V. CLASS III: COALITIONAL GRAPH GAMES

### A. Main Properties of Coalitional Graph Games

In canonical and coalition formation games, the utility or value of a coalition does *not* depend on how the players are interconnected within the coalition. However, it has been shown that, in certain scenarios, the *underlying communication structure* between the players in a coalitional game can have a major impact on the utility and other characteristics of the game [14], [34]. By the underlying communication structure, we mean the graph representing the connectivity of the players among each other, i.e., which player communicates with which one inside each and every coalition. We illustrated examples on such interconnections in Section II and Fig. 2 (b). In general, the main properties that distinguish a coalitional graph game are as follows:

1) The coalitional game is in graph form, and can be TU or NTU. However, the value of a coalition may depend on the external network structure as explained in Section II.
2) The *interconnection* between the players within each coalition, i.e., who is connected to whom, strongly impacts the characteristics and outcome of the game.
3) The main objective is to derive low complexity distributed algorithms for players that wish to build a network *graph* (directed or undirected) and not just coalitional groups as in coalition formation games (Class II). Another objective is to study the properties (stability, efficiency, etc) of the formed network graph.

In coalitional graph games, the main theme is the presence of a graph for communication between the players. Typically, there are two objectives for coalitional graph games. The first and most important objective, is to provide low complexity algorithms for building a network graph to connect the players. A second objective is to study the properties and stability of the formed network graph. In some scenarios, the network graph is given, and hence analyzing its stability and efficiency is the only goal of the game. The following sections provide an in-depth study of coalitional graph games.

### B. Coalitional Graph Games and Network Formation Games

The idea of having a value dependent on a graph of communication between the players was first introduced by Myerson in [14], through the graph function for TU games. In this work, starting with a TU canonical coalitional game $(\mathcal{N}, v)$ and given an undirected graph $G$ that interconnects the players in the game, Myerson attempts to find a fair solution. For this purpose, a new value function $u$, which depends on the graph, is defined. For evaluating the value $u$ of a coalition $S$, this coalition is divided into smaller coalitions depending on the players that are connected through $S$. For example, given a 3-players coalition $S = \{1, 2, 3\}$ and a graph $G = \{(2, 3)\}$ (only players 2 and 3 are connected by a link in $G$), the value $u(S, G)$ is equal to $u(S, G) = v(\{2, 3\}) + v(\{1\})$, where $v$ is the original value of the canonical game. Using the new value $u$, Myerson presents an axiomatic approach, similar to the Shapley value, for solving the game in graph function form. The work in [14] shows that, a fair solution of the canonical game $(\mathcal{N}, v)$ in the presence of a graph structure, is the Shapley value of the game $(\mathcal{N}, u)$ where $u$ is the newly defined value. This solution is known as the *Myerson value*. The drawback of the approach in [14] is that the value $u$ of a coalition depends only on the connected players in the coalition with no dependence on the structure, e.g., for both graphs $G_S^1$ and $G_S^2$ in Fig. 2 (b), the values $u$ are equal (although the payoffs received by the players in $G_S^1$ and $G_S^2$ through the Myerson value allocation would be different due to the different graphs).

Nevertheless, the work in [14] motivated future work, and in [15] the value was extended so as to depend on the graph structure, and not only on the connected components. By doing so, coalitional graph games became a richer framework, however, finding solutions became more complex. While in [14], the objective was to find a solution, *given a graph*, new research in the area sought algorithms for *forming the graph*. One prominent tool in this area is non-cooperative game theory which was extensively used for forming the network graph. For instance, in [1, Ch. 9.5], using the Myerson framework of [14], an extensive form game is proposed for forming the network graph. However, the extensive form approach is impractical in many situations, as it requires listing all possible links in the graph, which is a complex combinatorial problem. Nonetheless, a new breed of games started to appear following this work, and these games are known as *network formation games*. The main objective in these games is to study the interactions among a group of players that wish to form a graph. Although in some references these games are decoupled from coalitional game theory, we place these games under coalitional graph games due to several reasons: (i)- The basis of all network formation games is the work on coalitional graph games that started in [14], (ii)- network formation games share many objectives with coalitional graph games such as the presence of a value and an allocation rule, the need for stability among others, and (iii)- the solutions of network formation games are quite correlated with coalition formation games (in terms of forming the graph) and canonical games (in terms of having stable allocations).

Network formation games can be thought of as a hybrid between coalitional graph games and non-cooperative games. The reason is that, for forming the network, non-cooperative game theory plays a prominent role. Hence, in network formation games there is a need to form a network graph as well as to ensure the stability of this graph, through concepts analogous to those used in canonical coalitional games. For forming the graph, a broad range of approaches exist, and are grouped into two types: *myopic* and *far sighted* [7]. The main difference between these two types is that, in myopic approaches, the players play their strategies given the current state of the network, while in far sighted algorithms, the players adapt their strategy by learning, and predicting future strategies of the other players. For both approaches, well-known concepts from non-cooperative game theory can be used. The most popular of such approaches is to consider the network formation as a non-zero sum non-cooperative game, where the players' strategies are to select one or more links to form or break. One approach to solve the game is to play *myopic best response dynamics* whereby each player selects the strategy, i.e. the link(s) to form or break, that maximizes its utility. Under certain conditions on the utilities, the best response dynamics converge to a Nash equilibrium, which constitutes a Nash network. These approaches are widespread in network formation games [36–38], and also, several refinements to the Nash equilibrium suitable for network formation are used [36–38]. The main drawback of aiming for a Nash network is that, in many network formation games, the Nash networks are trivial graphs such as the empty graph or can be inefficient. For these reasons, a new type of network formation games has been developed, which utilizes new concepts for stability such as *pairwise stability* and *coalitional stability* [35]. The basic idea is to present stability notions that depend on deviations by a group of players instead of the unilateral deviations allowed by the Nash equilibrium. Independent of the stability concept, a key design issue in network formation games is the tradeoff between stability and efficiency. It is desirable to devise algorithms for forming stable networks that can also be efficient in terms of payoff distribution or total social welfare. Several approaches for devising such algorithms exist, notably using stochastic processes, graph theoretical techniques or non-cooperative games. For a comprehensive survey on such algorithms, we refer the reader to [35].

---

[7]These approaches are sometimes referred to as dynamics of network formation (see [35]).

Finally, the Myerson value and network formation games are not the only approaches for solving coalitional graph games. Other approaches, which are closely tied to canonical games can be proposed. For example, the work in [34], proposes to formulate a canonical game-like model for an NTU game, whereby the graph structure is taken into account. In this work, the authors propose an extension to the core called the *balanced core* which takes into account the graph structure. Further, under certain conditions, analogous to the balanced conditions of canonical games, the authors in [34] show that this balanced core is non-empty. Hence, coalitional graph games constitute quite a rich and diverse framework, with a broad range of applications. In the rest of this section, we review sample applications from communication networks.

*C. Applications of Coalitional Graph Games*

*1) Distributed uplink tree formation in IEEE 802.16j:* The most recent WiMAX standard, the IEEE 802.16j, introduced a new node, the relay station (RS) for improving the network's capacity and coverage. The introduction of the RS impacts the network architecture of WiMAX networks as the mesh network is replaced by a tree architecture which connects the base station (BS) to its subordinate RSs. An efficient design of the tree topology is, thus, a challenging problem, notably because the RSs can be nomadic or mobile. The IEEE 802.16j standard does not provide any algorithm for the tree formation, however, it states that both distributed and centralized approaches may be used. For tackling the design of the tree topology in 802.16j networks from a distributed approach, coalitional graph games provide a suitable framework. In [39], we model the problem of the *uplink* tree formation in 802.16j using coalitional graph games, namely network formation games. In this model, the players are the RSs who interact for forming a *directed* uplink tree structure (directed towards the BS). Every RS $i$ in the tree, acts as a source node, and transmits the packets that it receives from external mobile stations (MSs) to the BS, using multi-hop relaying. Hence, when RS $i$ is transmitting its data to the BS, all the RSs that are parents of $i$ in the tree relay the data of $i$ using decode-and-forward relaying. Through multi-hop relaying, the probability of error is reduced, and consequently the packet success rate (PSR) achieved by a RS can be improved. Essentially, the value function in this game is NTU as each RS optimizes its own utility. The utility of a RS $i$ is an increasing function of the effective number of packets received by the BS (effective throughput) while taking into account the PSR, as well as the number of packets received from other RS (the more a RS receives packet, the more it is rewarded by the network). The utility also reflects the cost of maintaining a link, hence, each RS $i$ has a maximum number of links that it can support. As the number of links on a RS $i$ increases, the rewards needed for *accepting* a link also increase, hence making it difficult for other RSs to form a link with $i$. The strategy of each RS is two-fold: (1)- Each RS can select another RS (or the BS) with whom to connect, and (2)- Each RS can choose to break a number of links that are connected to it. For forming a directed link $(i,j)$ between RS $i$ and RS $j$, the consent of RS $j$ is needed. In other words, if RS $i$ bids to connect to RS $j$, RS $j$ can either accept this link as a new connection, accept this link by replacing one or more other links, or reject the link. Using this formulation, the network formation game is a non-cooperative non-zero sum game played between the RSs, with the previously defined strategies. Hence, the dynamics of network formation are performed using an algorithm consisting of two phases. In the first phase, the RSs are prioritized, and in the second phase, proceeding sequentially by priority, each RS is allowed to play its best response, i.e., the strategy that maximizes its utility. This algorithm is myopic, since the best response of a RS is played given the current state of the network graph. The end result is the formation of a Nash network tree structure that links the RSs to the BS. This tree structure is shown in [39] to yield an improvement in the overall PSR achieved by the MSs in the network, compared to a static star topology or a network with no relays. The proposed algorithm allows each RS to autonomously choose whether to cooperate or not, and hence, it can easily be implemented in a distributed manner.

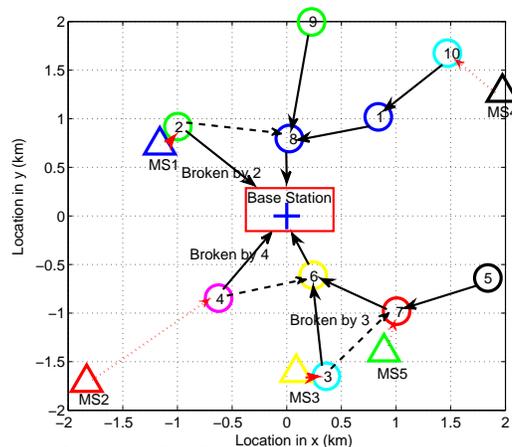

Fig. 5. Example of an 802.16j tree topology formed using a distributed network formation game as per [39].

In Fig. 5, we show an example of a network topology formed by 10 RSs. In this figure, the solid arrows indicate the network topology that formed before the deployment of any MSs (in the presence of keep-alive packets only). The proposed network formation algorithm is, in fact, adaptive to environmental changes, such as the deployment of the external MSs as well as mobility of the RSs or MSs. Hence, in Fig. 5, we can see how the RSs decide to break some of their link, replacing them with new links (in dashed arrows) hence adapting the topology, following the deployment of a number of MSs. In [40], the application of network formation games in 802.16j was extended and the algorithm was adapted to support the tradeoff between improving the effective throughput by relaying and the delay incurred by multi-hop transmission, for voice over IP services in particular. Future work can tackle various aspects of this problem using the tools of coalitional graph games. These aspects include devising a probabilistic approach to the network formation, or utilizing coalition graph games concepts such as the balanced core introduced in [34] among others.

*2) Other applications and future potential:* The presence of a network graph is ubiquitous in many wireless and communication applications. For designing, understanding, and analyzing such graphs, coalitional graph games are the accurate tool. Through the various concepts pertaining to network formation, stability, fairness, or others, one can model a diversity of problems. For instance, network formation games have been widely used in routing problems. For example, in [41], a stochastic approach for network formation is provided. In the proposed model, a network of nodes that are interested in forming a graph for routing traffic among themselves is considered. Each node in this model aims at minimizing its cost function which reflects the various costs that routing traffic can incur (routing cost, link maintenance cost, disconnection cost, etc.). For network formation, the work in [41] proposes a myopic dynamic best response algorithm. Each round of this algorithm begins by randomly selecting a pair of nodes $i$ and $j$ in the network. Once a random pair of nodes is selected, the algorithm proceeds in two steps. In the first step, if the link $(i,j)$ is already formed in the network, node $i$ is allowed to break this link, while in the second step node $i$ is allowed to form a new link with a certain node $k$, if $k$ accepts the formation of the link $(i,k)$. In the model of [41], the benefit from forming a link $(i,j)$ can be seen as some kind of cost sharing between nodes $i$ and $j$. By using a stochastic process approach, the work in [41] shows that the proposed myopic algorithm always converges to a pairwise stable and efficient tree network. Under a certain condition on the cost function, the stable and efficient tree network is a simple star network. The efficiency is measured in terms of Pareto optimality of the utilities as the proposed game is inherently NTU. Although the network formation algorithm in [41] converges to a stable and efficient network, it suffers from a major drawback which is the slow convergence time, notably for large networks. The proposed algorithm is mainly implemented for undirected graphs but the authors provide sufficient insights on how this work can extend to directed graphs.

The usage of network formation games in routing applications is not solely restricted to forming the network, but also for studying properties of an existing network. For instance, in [42], the authors study the stability and the flow of the traffic in a given directed graph. For this purpose, several concepts from network formation games such as pairwise stability are used. In addition, the work in [42] generalizes the concept of pairwise stability making it more suitable for directed graphs. Finally, [42] uses non-cooperative game theory to determine the network flows at different nodes while taking into account the stability of the network graph. The applications of coalitional graph games are by no means limited to routing problems. The main future potential of using this class of games lies in problems beyond network routing. For instance, coalitional graph games are suitable tools to analyze problems pertaining to information trust management in wireless networks, multi-hop cognitive radio, relay selection in cooperative communications, intrusion detection, peer-to-peer data transfer, multi-hop relaying, packet forwarding in sensor networks, and many others. Certainly, this rich framework is bound to be used thoroughly in the design of many aspects of future communication networks.

## VI. CONCLUSIONS

In this tutorial, we provided a comprehensive overview of coalitional game theory, and its usage in wireless and communication networks. For this purpose, we introduced a novel classification of coalitional games by grouping the sparse literature into three distinct classes of games: canonical coalitional games, coalition formation games, and coalitional graph games. For each class, we explained in details the fundamental properties, discussed the main solution concepts, and provided an in-depth analysis of the methodologies and approaches for using these games in both game theory and communication applications. The presented applications have been carefully selected from a broad range of areas spanning a diverse number of research problems. The tutorial also sheds light on future opportunities for using the strong analytical tool of coalitional games in a number of applications. In a nutshell, this article fills a void in existing communications literature, by providing a novel tutorial on applying coalitional game theory in communication networks through comprehensive theory and technical details as well as through practical examples drawn from both game theory and communication applications.

## REFERENCES

[1] R. B. Myerson, *Game Theory, Analysis of Conflict*. Cambridge, MA, USA: Harvard University Press, Sep. 1991.
[2] T. Başar and G. J. Olsder, *Dynamic Noncooperative Game Theory*. Philadelphia, PA, USA: SIAM Series in Classics in Applied Mathematics, Jan. 1999.
[3] G. Owen, *Game Theory, 3rd edition*. London, UK: Academic Press, Oct. 1995.
[4] Z. Han and K. J. Liu, *Resource Allocation for Wireless Networks: Basics, Techniques, and Applications*. New York, USA: Cambridge University Press, 2008.
[5] T. Alpcan and T. Başar, "A globally stable adaptive congestion control scheme for Internet-style networks with delay," *IEEE/ACM Trans. on Networking*, vol. 13, pp. 1261–1274, Dec. 2005.

[6] T. Alpçan, T. Başar, R. Srikant, and E. Altman, "CDMA uplink power control as a noncooperative game," *Wireless Networks*, vol. 8, pp. 659–670, 2002.
[7] A. MacKenzie, L. DaSilva, and W. Tranter, *Game Theory for Wireless Engineers*. Morgan and Claypool Publishers, Mar. 2006.
[8] T. Başar, "Control and game theoretic tools for communication networks (overview)," *App. Comput. Math.*, vol. 6, pp. 104–125, 2007.
[9] R. La and V. Anantharam, "A game-theoretic look at the Gaussian multiaccess channel," in *Proc. of the DIMACS Workshop on Network Information Theory*, New Jersey, NY, USA, Mar. 2003.
[10] S. Mathur, L. Sankaranarayanan, and N. Mandayam, "Coalitions in cooperative wireless networks," *IEEE J. Select. Areas Commun.*, vol. 26, pp. 1104–1115, Sep. 2008.
[11] J. von Neumann and O. Morgenstern, *Theory of Games and Economic Behavior*. Princeton, NJ, USA: Princeton University Press, Sep. 1944.
[12] R. J. Aumann and B. Peleg, "Von neumann-morgenstern solutions to cooperative games without side payments," *Bulletin of American Mathematical Society*, vol. 6, pp. 173–179, 1960.
[13] R. Thrall and W. Lucas, "N-person games in partition function form," *Naval Research Logistics Quarterly*, vol. 10, pp. 281–298, 1963.
[14] R. Myerson, "Graphs and cooperation in games," *Mathematics of Operations Research*, vol. 2, pp. 225–229, Jun. 1977.
[15] M. Jackson and A. Wolinsky, "A strategic model of social and economic networks," *Journal of Economic Theory*, vol. 71, pp. 44–74, 1996.
[16] V. Conitzer and T. Sandholm, "Complexity of determining nonemptiness of the core," CMU, Tech. Rep. CS-02-137.
[17] M. Madiman, "Cores of cooperative games in information theory," *EURASIP Journal on Wireless Communications and Networking*, vol. 2008, Mar. 2008.
[18] A. Aram, C. Singh, and S. Sarkar, "Cooperative profit sharing in coalition based resource allocation in wireless networks," in *Proc. of IEEE INFOCOM*, Rio de Janeiro, Brazil, Apr. 2009.
[19] Z. Han and V. Poor, "Coalition games with cooperative transmission: a cure for the curse of boundary nodes in selfish packet-forwarding wireless networks," *IEEE Trans. Comm.*, vol. 57, pp. 203–213, Jan. 2009.
[20] L. Shapley and M. Shubik, "The assignment game i: The core," *Int. Journal of Game Theory*, vol. 1, pp. 111–130, 1972.
[21] J. Cai and U. Pooch, "Allocate fair payoff for cooperation in wireless ad hoc networks using shapley value," in *Proc. Int. Parallel and Distributed Processing Symposium*, Santa Fe, NM, USA, Apr. 2004, pp. 219–227.
[22] J. Castro, D. Gomez, and J. Tejada, "Polynomial calculation of the shapley value based on sampling," *Computers and Operations Research*, vol. 36, pp. 1726–1730, May 2009.
[23] J. Derks and J. Kuipers, "Implementing the simplex method for computing the prenucleolus of transferable utility games," University of Maastricht, the Netherlands, Tech. Rep., 1997.
[24] R. Aumann and J. Drèze, "Cooperative games with coalition structures," *Int. Journal of Game Theory*, vol. 3, pp. 317 – 237, Dec. 1974.
[25] T. Sandholm, K. Larson, M. Anderson, O. Shehory, and F. Tohme, "Coalition structure generation with worst case guarantees," *Artifical Intelligence*, vol. 10, pp. 209–238, Jul. 1999.
[26] D. Ray, *A Game-Theoretic Perspective on Coalition Formation*. New York, USA: Oxford University Press, Jan. 2007.
[27] K. Apt and A. Witzel, "A generic approach to coalition formation," in *Proc. of the Int. Workshop on Computational Social Choice (COMSOC)*, Amsterdam, the Netherlands, Dec. 2006.
[28] T. Arnold and U. Schwalbe, "Dynamic coalition formation and the core," *Journal of Economic Behavior and Organization*, vol. 49, pp. 363–380, Nov. 2002.
[29] A. Bogomonlaia and M. Jackson, "The stability of hedonic coalition structures," *Games and Economic Behavior*, vol. 38, pp. 201–230, Jan. 2002.
[30] W. Saad, Z. Han, T. Başar, M. Debbah, and A. Hjørungnes, "A selfish approach to coalition formation among unmanned aerial vehices in wireless networks," in *Proc. Int. Conf. on Game Theory for Networks*, Istanbul, Turkey, May 2009.
[31] W. Saad, Z. Han, M. Debbah, and A. Hjørungnes, "A distributed merge and split algorithm for fair cooperation in wireless networks," in *Proc. Int. Conf. on Communications, Workshop on Cooperative Communications and Networking*, Beijing, China, May 2008.
[32] W. Saad, Z. Han, M. Debbah, A. Hjørungnes, and T. Başar, "Coalitional games for distributed collaborative spectrum sensing in cognitive radio networks," in *Proc. of IEEE INFOCOM*, Rio de Janeiro, Brazil, Apr. 2009.
[33] W. Saad, Z. Han, T. Başar, M. Debbah, and A. Hjørungnes, "Physical layer security: Coalitional games for distributed cooperation," in *Proc. 7th Int. Symp. on Modeling and Optimization in Mobile, Ad Hoc, and Wireless Networks (WiOpt)*, Seoul, South Korea, Jun. 2009.
[34] P. Herings, G. van der Laan, and D. Talman, "Cooperative games in graph structure," *Research Memoranda, Maastricht Research School of Economics, of Technology and Organization*, no. 11, Aug. 2002.
[35] M. O. Jackson, "A survey of models of network formation: Stability and efficiency," *Working Paper 1161, California Institute of Technology, Division of the Humanities and Social Sciences*, Nov. 2003.
[36] G. Demange and M. Wooders, *Group Formation in Economics*. New York, USA: Cambridge University Press, 2006.
[37] V. Bala and S. Goyal, "Noncooperative model of network formation," *Econometrica*, vol. 68, pp. 1181–1230, Sep. 2000.
[38] J. Derks, J. Kuipers, M. Tennekes, and F. Thuijsman, "Local dynamics in network formation," in *Proc. Third World Congress of The Game Theory Society*, Illinois, USA, Jul. 2008.
[39] W. Saad, Z. Han, M. Debbah, and A. Hjørungnes, "Network formation games for distributed uplink tree construction in IEEE 802.16j networks," in *Proc. IEEE Global Communication Conference*, New Orleans, LA, USA, Dec. 2008.
[40] W. Saad, Z. Han, M. Debbah, A. Hjørungnes, and T. Başar, "A game-based self-organizing uplink tree for voip services in IEEE 802.16j networks," in *Proc. Int. Conf. on Communications*, Dresden, Germany, Jun. 2009.
[41] E. Arcaute, R. Johari, and S. Mannor, "Network formation: Bilateral contracting and myopic dynamics," *Lecture Notes in Computer Science*, vol. 4858, pp. 191–207, Dec. 2007.
[42] R. Johari, S. Mannor, and J. Tsitsiklis, "A contract based model for directed network formation," *Games and Economic Behavior*, vol. 56, pp. 201–224, Jul. 2006.